\documentclass[]{spie}  

\usepackage{amsmath,amsfonts,amssymb}
\usepackage{graphicx}
\usepackage[colorlinks=true, allcolors=blue]{hyperref}

\title{Design and development of a high-speed Visible Pyramid Wavefront Sensor for the MMT AO system}

\author[a,*]{Narsireddy Anugu}
\author[a]{Olivier Durney }
\author[a]{Katie M. Morzinski}
\author[b]{Phil Hinz}
\author[c]{Suresh Sivanandam }
\author[a]{Jared Males }
\author[a]{Andrew Gardner}
\author[d]{Chuck Fellows }
\author[a]{Manny Montoya }
\author[a]{Grant West }
\author[a]{Amali Vaz }
\author[a]{Emily Mailhot}
\author[a]{Jared Carlson}
\author[c]{Shaojie Chen }
\author[c]{Masen Lamb }
\author[c]{Adam Butko }
\author[a]{Elwood Downey }
\author[c]{Jacob Tylor }
\author[a]{Buell Jannuzi }

\affil[a]{Steward Observatory, University of Arizona, 933 N. Cherry Ave., Tucson AZ 85721, USA}
\affil[b]{University of California, Santa Cruz 1156 High St, Santa Cruz, CA 95064}
\affil[c]{Department of Astronomy and Astrophysics, University of Toronto, 50 St. George Street, Toronto, ON, Canada, M5S 3H4}
\affil[d]{Lunar and Planetary Laboratory, University of Arizona, 1629 E University Blvd, Tucson, AZ 85721}

\authorinfo{* Steward Observatory Fellow in Astronomical Instrumentation and
Technology. \\
E-mail: \href{nanugu@arizona.edu}{nanugu@arizona.edu}}

\begin{document} 
\maketitle

\begin{abstract}
MAPS, MMT Adaptive optics exoPlanet characterization System, is the upgrade of legacy 6.5m MMT adaptive optics system. It is an NSF MSIP-funded project that includes (i) refurbishing of the MMT Adaptive Secondary Mirror (ASM), (ii) new high sensitive and high spatial order visible and near-infrared pyramid wavefront sensors, and (iii) the upgrade of Arizona Infrared Imager and Echelle Spectrograph (ARIES) and MMT high Precision Imaging Polarimeter (MMTPol) science cameras. This paper will present the design and development of the visible pyramid wavefront sensor. This system consists of an acquisition camera, a fast-steering tip-tilt modulation mirror, a double pyramid, a pupil imaging triplet lens, and a low noise and high-speed frame rate based CCID75 camera. We will report on hardware and software and present the laboratory characterization results of the individual subsystems, and outline the on-sky commissioning plan.
\end{abstract}



\section{INTRODUCTION}\label{sec:intro}  
MAPS\cite{Morzinski2020}, MMT Adaptive Optics exoPlanet characterization System is the upgrade of  MMT 6.5m\cite{Williams2018SPIE10700E..2TW}  20-year old adaptive optics system\cite{Lloyd-Hart2000SPIE.4007..167L,Brusa2003SPIE.4839..691B}. MAPS is an NSF MSIP-funded project (PI: Phil Hinz; now -- Katie M. Morzinski) that includes (i) development of new actuators and control electronics for the adaptive-secondary-mirror (ASM)\cite{Hinz2018},  (ii) new high-sensitive and high-spatial-order visible and near-infrared pyramid wavefront sensors (VPyWFS and IRPyWFS), and (iii) the upgrade of science cameras: Arizona infrared imager and echelle spectrograph (ARIES)\cite{McCarthy1998SPIE.3354..750M}, and MMT High Precision Imaging Polarimeter (MMTPol)\cite{Jones2007AAS...211.1124J}. The key advantage of having of ASM type deformable mirror is that it reduces the number of optical elements in the telescope’s AO system, thereby decreases the thermal noise of the system ($<10$\% emissivity\cite{Lloyd-Hart2000PASP..112..264L}) and increases throughput.  The MAPS key astrophysical science programs\cite{Morzinski2020} are to study exoplanet atmospheres\cite{Birkby2017AJ....153..138B} and debris disks in the near-infrared and thermal wavelengths ranging 1--5 $\mu$m by exploiting the MMT 6.5m collecting area and its high-angular resolutions, and ARIES high-spectral resolutions  and MMTPol precision polarization measurements. MAPS is a collaborative project with research groups from the University of Arizona, Arizona State University, Minnesota Institute for Astrophysics, and the University of Toronto. 


\begin{figure}[ht]
\centering
\includegraphics[width=0.7\textwidth]{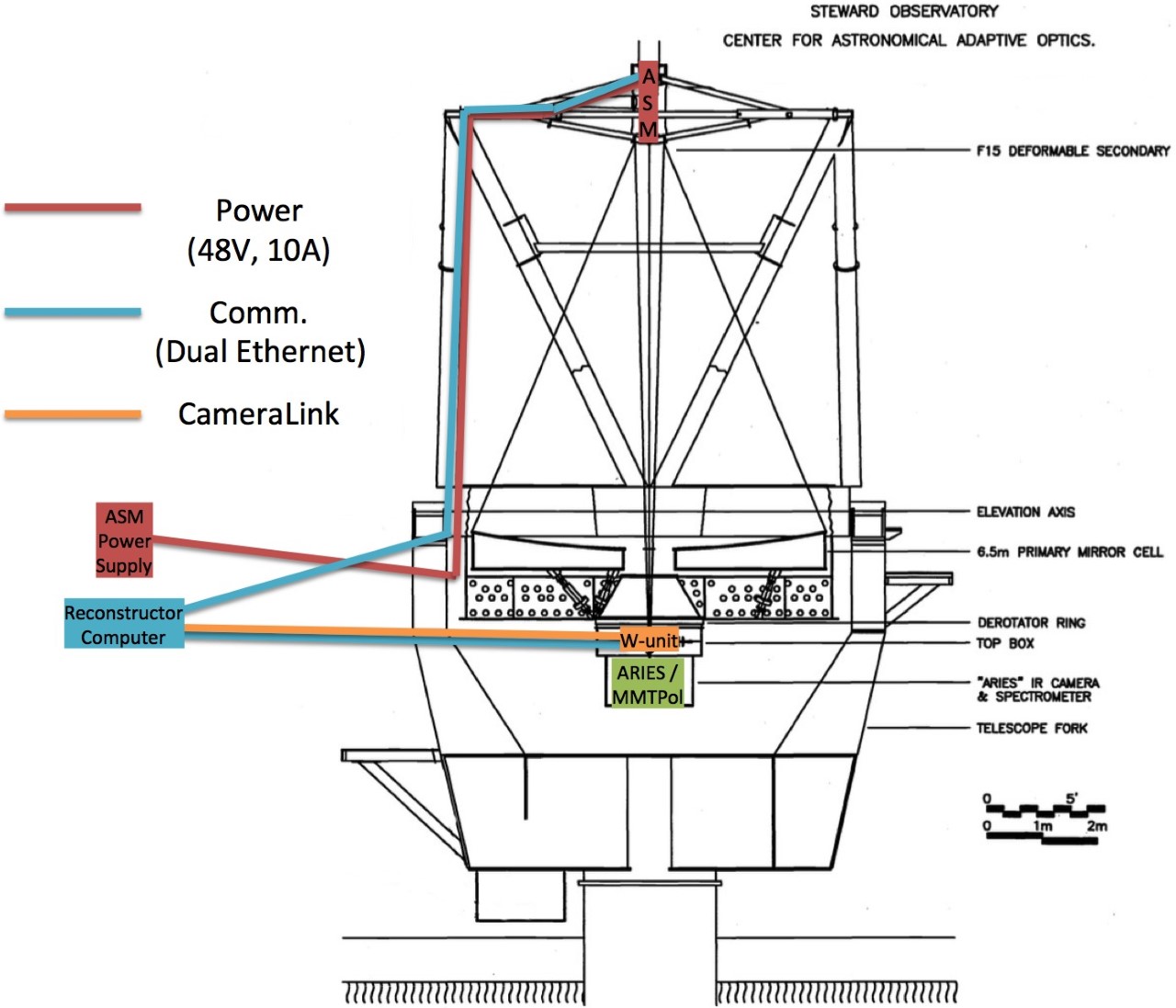}
\caption[example]{\label{MAPS overview} MAPS on the MMT~6.5m - showing ASM, wavefront sensors unit (W-unit) and science cameras: ARIES and MMTPol. The pyramid wavefront sensor images are read to \texttt{aocameras} computer using a camera link. The computed slopes are sent to AO Reconstructor Computer (\texttt{aorc}). The \texttt{aorc} computes the ASM position control commands from the slopes matrix and sends them to the ASM central hub for making corrections. Next ARIES and MMTPol records AO corrected science observations.}
\end{figure}

\begin{figure}
\centering
\includegraphics[width=0.9\textwidth]{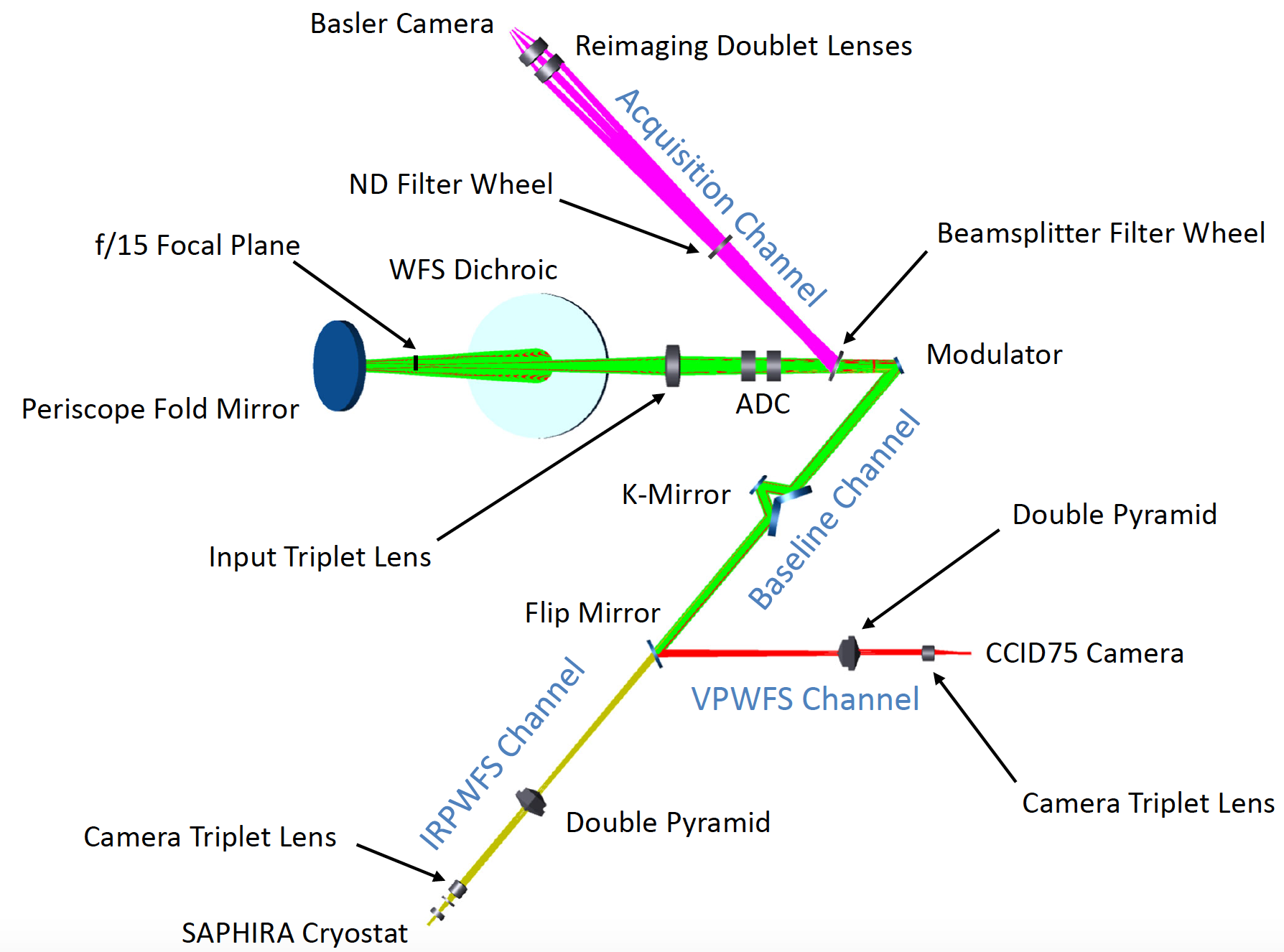}\\
\includegraphics[width=0.9\textwidth]{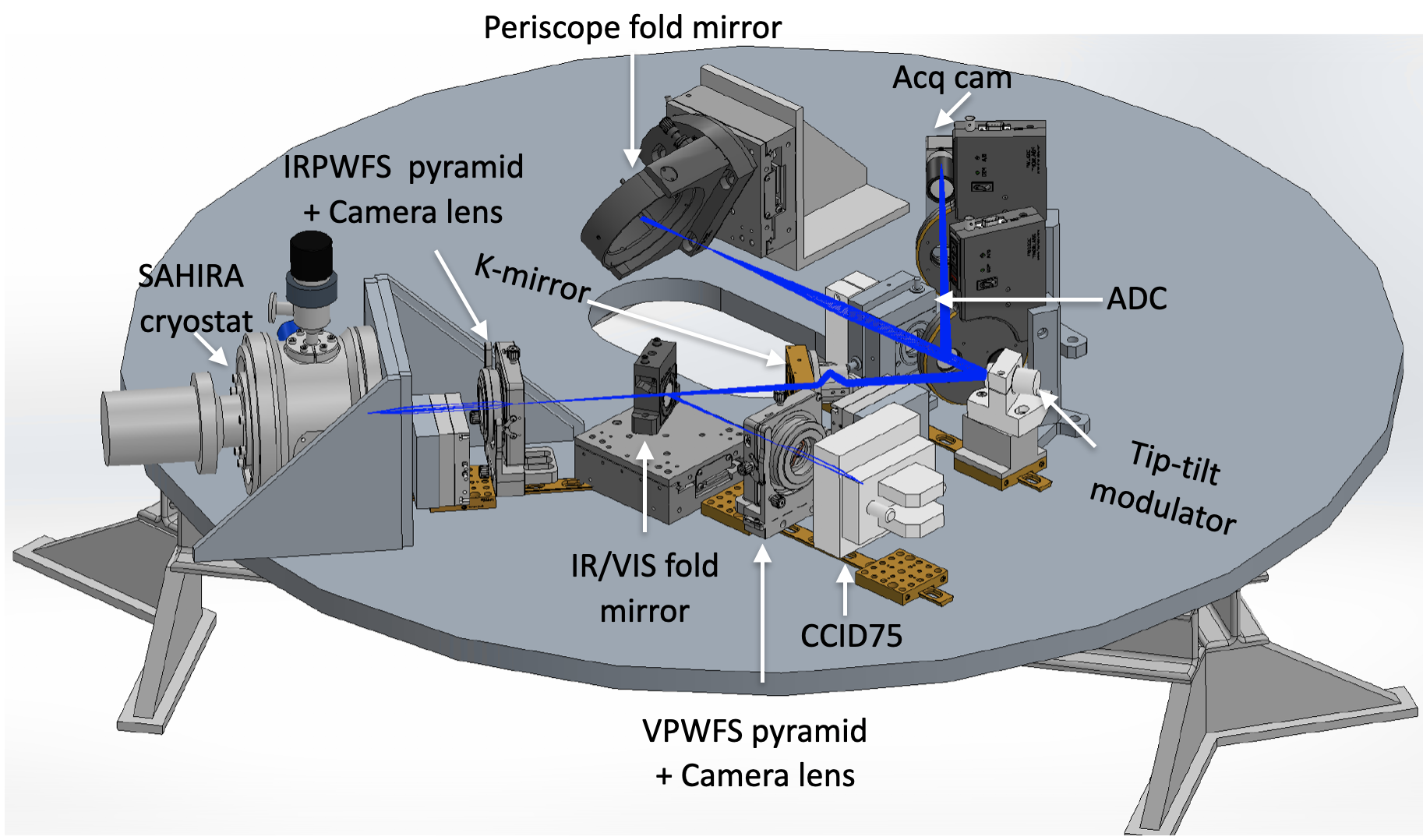}
\caption[example]{\label{design} (Top) optical and (bottom) mechanical design of MAPS waterfront sensor unit.}
\end{figure}

An overall design of the MAPS-ASM (see Hinz et al. 2018\cite{Hinz2018}) and infrared pyramid wavefront sensor (see Liu et al. 2018\cite{Liu2018} and Butko et al. 2018\cite{Butko2018SPIE10702E..3NB}) are reported in the previous SPIE proceedings. New actuators and control electronics for the ASM have been built and characterized thoroughly at the labs of Steward Observatory, University of Arizona. The ASM performance results (Vaz et al. 2020\cite{Vaz2020}) and a general overview of MAPS (Morzinski et al. 2020\cite{Morzinski2020}) can be found in this proceedings.  

Figure~\ref{MAPS overview} shows the layout of the MAPS on MMT~6.5m when installed. The VPyWFS and IRPyWFS are in the post-final-design-review stage, with their review completed in July 2020 (committee: Dr. Brian McLeod, Dr. Sandrine Thomas, and Dr. Jared Males). This paper will present the design and development of VPyWFS.  Section~\ref{optics-design} presents the opto-mechanical design of VPyWFS, which consists of an atmospheric dispersion corrector (ADC),  a K-mirror,  a fast-steering tip-tilt modulation mirror, an acquisition camera, a double pyramid sensor, a pupil imaging triplet lens, and a low readout noise and high-speed frame rate camera. We already acquired critical subsystems of VPyWFS, including a wavefront sensor camera, fast steering tip-tilt modulator, and acquisition camera. Their performance is characterized and is summarized in Section~\ref{Characterization}. In the end, we outline the current status of the project and on-sky commissioning plans.

\section{Design of  Visible wavefront sensor}
\subsection{Opto mechanical design}\label{optics-design}
VPyWFS is designed to $30\times30$ sub-aperture wavefront sensing with 1ms temporal response, and we plan to correct 220 modes of the adaptive secondary mirror. We have completed all the optical (lead: Olivier Durney) and mechanical (lead: Manny Montoya) design. Procurement of all parts requested. The design and component specifications have been carefully developed to ensure that we meet the very demanding performance requirements for studying exoplanets systems\cite{Morzinski2020}.   Figure~\ref{design} shows the high-level architecture of VPyWFS. The VPyWFS system is divided into four channels: 

\begin{itemize}
    \item \textbf{Base channel} consists of a WFS dichroic, periscope fold mirror, input triplet lens,  atmospheric dispersion corrector, beamsplitter, fast steering tip-tilt modulator, and image derotator (K-mirror).

    \item \textbf{Acquisition channel} consists of a Neutral Density (ND) filter wheel, imaging doublet lenses, a Basler acquisition camera.

    \item \textbf{VPyWFS Channel} consists of a double pyramid, pupil imaging camera triplet lens and CCID75 camera.

    \item \textbf{IRPyWFS Channel} consists of a similar double pyramid, pupil imaging camera triplet lens, and SAPHIRA camera in a cryostat.
\end{itemize}

\subsubsection{The base channel}
The  MMT 6.5~m primary mirror and 0.642~m ASM create an f/15 converging beam, and this beam is the input for the WFS unit. The WFS dichroic splits the light between the WFS and ARIES or MMTPol science camera. An advanced custom-designed periscope fold mirror is designed to enable (i) 1~arc-minute WFS patrol field to select AO natural guide star and (ii) WFS nodding to take sky-backgrounds on the science cameras. The one arc-minute patrolling field capability increases the availability of AO natural guide stars.

\textbf{Input triplet} is a custom designed 35~mm lens that re‐images the input telescope f/15 beam onto the tip of the double pyramid of either VPyWFS or IRPyWFS with f/50 focus. Glass substrates used in this triplet are S‐FPL53, S‐NSL5, and S‐TIL1. The input triplet is on procurement from Rainbow optics.

\textbf{Atmospheric Dispersion Corrector (ADC)} is used to correct PSF elongation caused due to the atmospheric dispersion for the operational waveband of the wavefront sensors $0.6-1.8~\mu$m. For example, the atmospheric dispersion in the visible band ($0.6-1.0~\mu$m) causes an elongation of PSF around 740~mas for the Zenith angle of $60^\circ$. In contrast, the diffraction-limited PSF size in visible is 19-31 mas. The mitigation of this dispersion is essential for wavefront sensing. A custom ADC is designed using two prisms with a size of 25.4mm separated by 9mm, and with prism angle of $74.741^\circ$. Glass substrates used for the prisms are S‐PHM53 and S‐TIM8. This is being procured from BMV Optical Technologies. 

\textbf{Beamsplitter filter wheel} splits the light and shares between the acquisition and wavefront sensor channels. 

\textbf{A fast steering tip-tilt modulation mirror} is used to increase  dynamic range of our pyramid wavefront sensor because the sensor quickly becomes non-linear and can saturate with small wavefront aberrations. A circular modulation is selected to use with modulation amplitude of $\pm 10 \lambda/D$, i.e., $\pm714~\mu$rad at $\lambda=700$nm to account for worse atmospheric turbulence. The pyramid modulation is conducted with high dynamic and stiffness Piezo tip-tilt platform assembly, PI S‐331.2SH. This tip-tilt stage has strain gauges to measure the position of the mirror moving. A 15mm silver coated mirror from Edmund Optics, 34-388, is used. A circular modulation is generated by applying a sine waveform on the two tip-tilt axes with a 90-degree shift between them.   This tip-tilt stage is controlled by an E-727 digital three-channel Piezo controller.

\textbf{K‐Mirror} is an image derotator to keep the pupil aligned between ASM and VPyWFS or IRPyWFS to account for the Earth's rotation. We made a custom design that consists of a prism and a mirror. A prism is used instead of individual mirrors K‐M1 and K‐M3. The prism facets are protected silver coated and have sizes of $31 \times 22$~mm. A 20~mm mirror (K‐M2) is mounted on a 3‐degrees of freedom stage to co‐align input and output beams to the optical axis. The K-mirror is being procured from BMV Optical Technologies. 

\begin{figure}[ht]
\centering
\includegraphics[width=\textwidth]{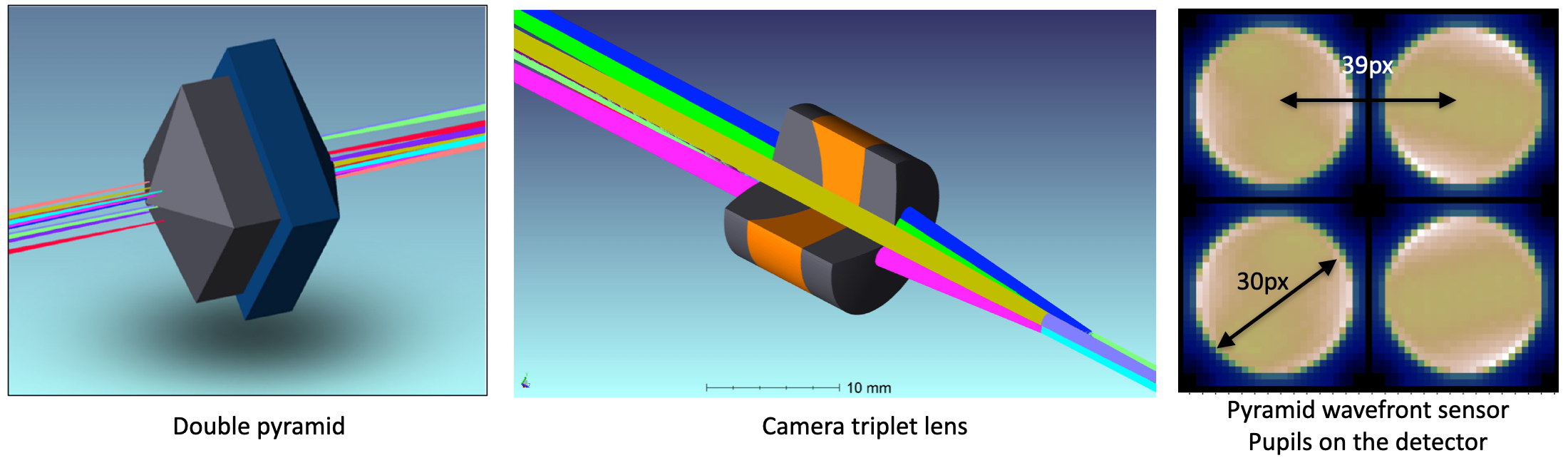}
\caption[example]{\label{VPyWFS} The visible pyramid wavefront sensor: (left) double pyramid, (middle) camera triplet lens and (right) Yorick-Yao\cite{Rigaut2013aoel.confE..18R} simulated pyramid wavefront sensor pupil images on the detector.}
\end{figure}

\subsubsection{The VPyWFS channel}
A pyramid wavefront sensor is a pupil plane wavefront sensor that splits the focal plane into four quadrants on a detector\cite{Ragazzoni1996JMOp...43..289R}. The wavefront slopes from the detector images are calculated using the quad-cell centroid of pixel intensities. The number of sub-apertures in a classical Shack-Hartmann lenslet array is equal to the number of pixels across one of the pupils in the pyramid wavefront sensor. The pyramid wavefront sensor is very sensitive\cite{Ragazzoni1999A&A...350L..23R} compared to the Shack-Hartmann sensor by using less number of pixels and also having the freedom of changing the dynamic range by adjusting the radius of modulation up to $\pm10\lambda/D$ for bright stars and/or binning the detector camera pixels for faint stars.  

The VPyWFS system operates in $0.6-1\mu$m wavelengths. Its design (see Figure~\ref{VPyWFS}) consists of an achromatic double pyramid, a pupil imaging triplet lens, and the CCID75 detector camera. Table~\ref{VPyWFS_spec} summarize the design specifications.

\textbf{Double pyramid} splits the focal plane into four quadrants. It consists of two four-sided prisms aligned back to back. The Zemax design is taken from the LBTI's design -- a highly successful pyramid wavefront sensor\cite{Tozzi2008SPIE.7015E..58T} -- and optimized for MAPS specifications. For easy manufacturing, the form factor was changed from round to square. Each pyramid is a 4‐faceted isosceles pyramid with an extruded square base. Critical pyramid parameters like apex angles and center thicknesses are optimized for the $1/10$ pixel shift/blur requirements over a range of modulations from none to $10\lambda/D$. The front prism $16 \times 16$~mm is made from  N-SK11, and the back prism $20 \times 20$~mm is made from  N-PSK53, and they are bonded back‐to‐back using Norland61. The double pyramid is being on procurement from WZWOPTICAG.

\textbf{Camera triplet lens:}  The pyramid wavefront sensor's success depends on the size and separation of the pupils on the detector. We selected $30\times30$ pixels pupil size for the high-spatial orders of wavefront measurement. The pupil separation between any two pupil quadrants is $39$ pixels. The 39-pixel pupil separation is problematic for binning (factor 2 and 4), and so we plan to defocus and increase this to 40 pixels. A custom achromatic triplet was designed to re-image the four quadrants of the pupil plane on our CCID75 camera\cite{Schuette2014} with the appropriate pupil size and separation geometry. Glass substrates used are N‐BAK1, N‐KZFS5, and S‐LAM60. We require $80\times80$ pixels Regions Of Interest (ROI) on the detector. We are procuring it from Rainbow Research optics.

\begin{table}
\caption{The visible pyramid wavefront sensor design specifications.} 
\label{VPyWFS_spec}
\begin{center}       
\begin{tabular}{ll}
\hline
Parameter & value \\
\hline
Pupil diameter (sub-ap) & 30 px\\
Final pupil separation & 39 px (we aim 40 px)\\
Wavelength range & $0.6-1.0\mu$m\\
Field of view & $2\times2^{\prime\prime}$\\
Modulation amplitude & $\leq \pm10\lambda/D$\\
Pupil Jitter [pixel] & $\leq 1/10$ px\\
Pupil distortion & $\leq 1/10$ px\\
Detector size & CCID75, $160\times160$  ${\rm px}^2$\\
Detector pixel pitch & $21\mu$m\\
Detector ROI & $80\times80$ ${\rm px}^2$\\
Input F/\# & f/15\\
Frame rate & 1kHz\\
\hline
\end{tabular}
\end{center}
\end{table}

\begin{figure}[ht]
\centering
\includegraphics[width=\textwidth]{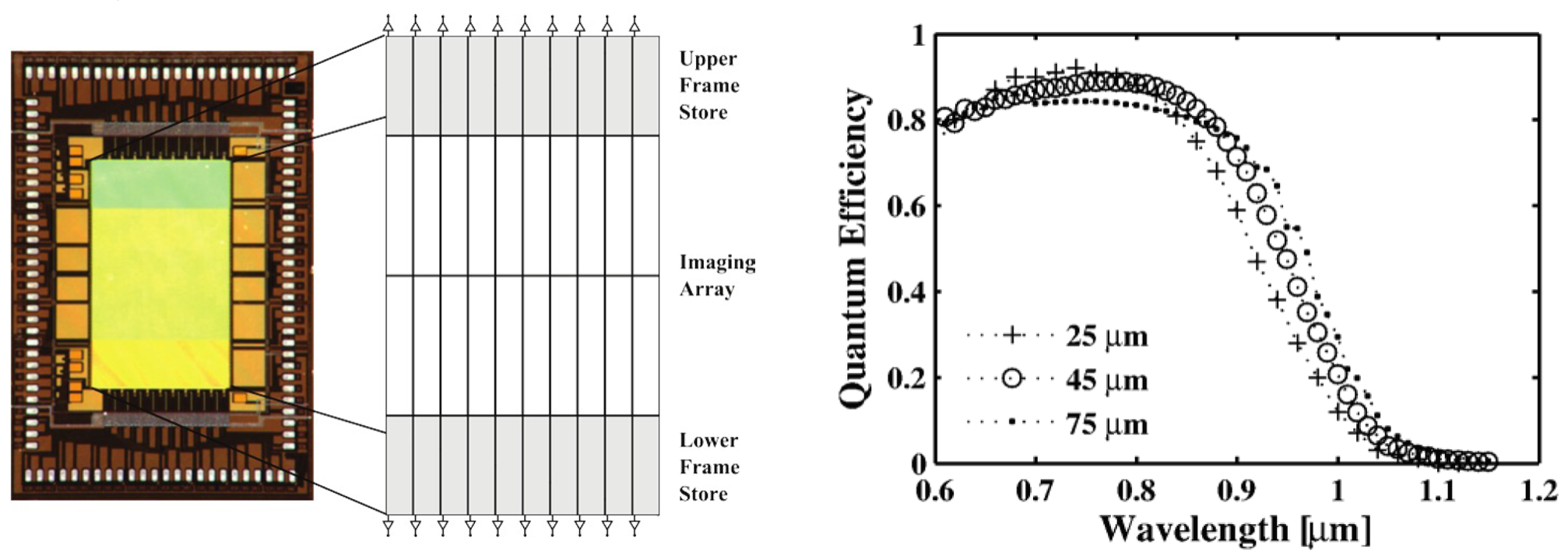}
\caption[example]{\label{ccid_75} (Left) The architecture of CCID75 reproduced from Schuette et al. (2014)\cite{Schuette2014}. It has 160 (Horizontal) $\times 160$ (Vertical) imaging area -- subdivided into twenty 16 (H) $\times 80$ (V) pixel channels. Each channel has a separate readout chain (serial readout register and output amplifier). (Right) Quantum efficiency is around 80\% for $0.6-0.9\mu$m wavelengths.}
\end{figure}

\textbf{CCID75}\cite{Schuette2014} is a $160\times160$-pixel CCD that has been demonstrated to operate at frame rates above 1 kHz with readout noise $\sigma_{\rm R}<3$e-/px. This camera does not use electron avalanche photodiode (eAPD) technology as in electron-multiplying CCD (EMCCD) but still provides arguably a low readout noise and fast frame rates using deeply depleted split frame transfer and 20 readout outputs.  A drawback of EMCCD is that the avalanche gain does not come for free but with excess noise, typically 1.47, as the avalanche-multiplication is statistical in nature, leading to increased noise in the output signal. Furthermore, over usage of the avalanche gain attracts trap assisted tunneling noise\cite{Finger2014SPIE.9148E..17F}. EMCCD is the best suited for applications where the signal levels are readout noise limited, and the conventional CCD output performs when available signal levels are photon-noise limited. 

Figure~\ref{ccid_75} shows the CCID75 architecture, (i) a deeply depleted split frame transfer CCD with two frame store areas, and (ii) 20 readout outputs. Each output is associated with an array of 16 (Horizontal) $\times 80$ (Vertical) imaging pixels. The CCID75 camera is provided courtesy of Air Force Research Laboratory and MIT Lincoln Laboratory. SciMeasure built the CCID75  control electronics. For this camera, the SciMeasure built 32 programs, called ReCall Settings (\texttt{RCL}) in the non-volatile memory of the controller. By selecting an appropriate \texttt{RCL}, we can get appropriate ROI, on-chip binning, and frame rate.  The characterization details of CCID75 are found in Section~\ref{Characterization}.

\subsubsection{IRPyWFS channel} The availability of two wavefront sensors covering the visible and near-infrared wavelengths is rare, but it allows a 100\% efficiency by boosting the sky coverage of AO guide stars. IRPyWFS would allow the use of the science object as the guide star. IRPyWFS will operate in 0.85-1.8$\mu$m, allowing red type stars, including M-type, to be used for AO guidance.  IRPyWFS uses a breakthrough sub-electron and fast frame rate detector based on electron avalanche photodiode (eAPD) detector, SAPHIRA\cite{Finger2014SPIE.9148E..17F}. The details of the design,  cryostat, and detector are described in previous SPIE proceedings\cite{Liu2018, Butko2018SPIE10702E..3NB}. A flip mirror is used to select VPyWFS or IRPyWFS based on the science requirement and color and brightness of the guide star in the field of view (see Figure~\ref{design}). 

\subsubsection{Acquisition camera channel}
\begin{figure}[ht]
\centering
\includegraphics[width=0.8\textwidth]{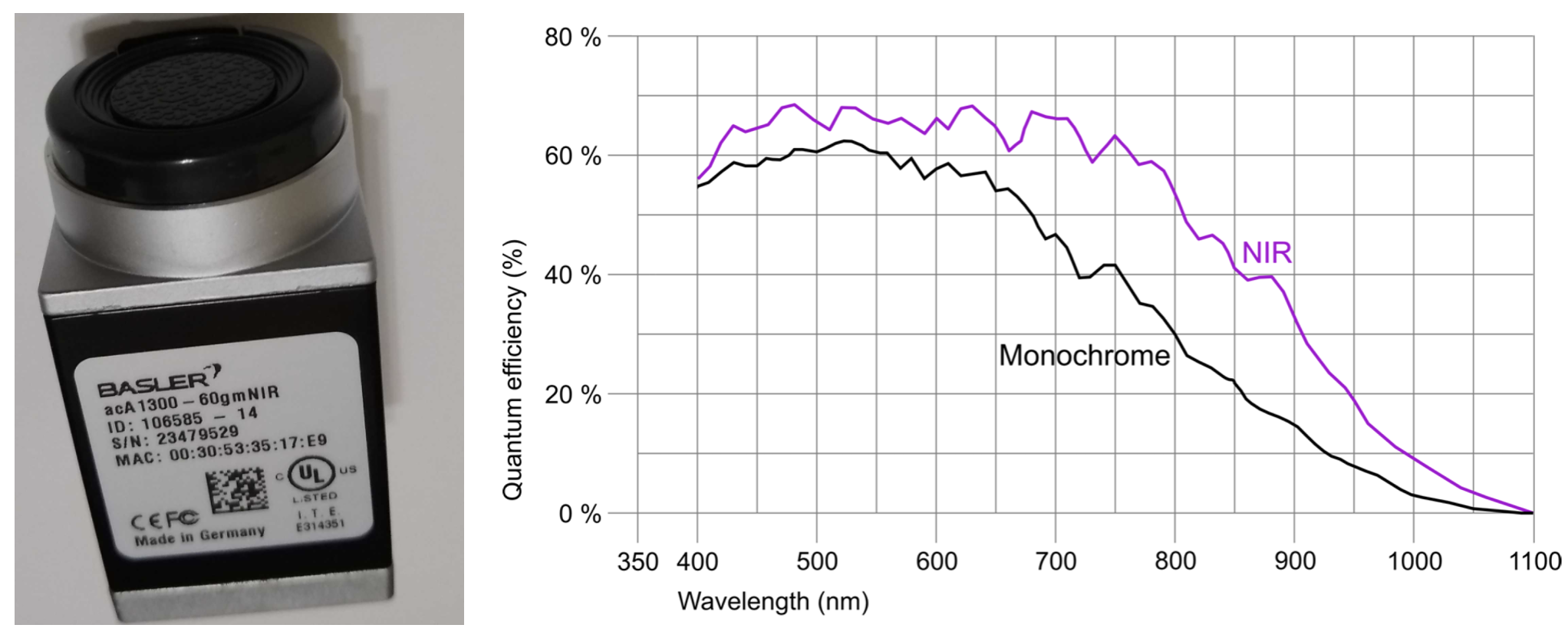}
\caption[example]{\label{acq}(Left) Acquisition camera procured from Basler. (Right) Quantum efficiency more than 65\% for wavelengths $0.4-0.8\mu$m.}
\end{figure}

The acquisition camera operates in $0.65 - 1.05 \mu$m wavelengths, and it enables us to find the target and AO guide stars in the field of view. A custom double achromatic doublet lens separated by 7 mm with a diameter 25.4 mm is used to image the incoming beam on a Baseler camera, CABR720 (see Figure~\ref{acq}). This lens will be attached to the camera in C‐mount.  The Baseler CABR720 camera is a $1280 \times 1024$ NIR CMOS GigE camera with $5.3\mu$m pixel pitch, 60Hz frame rate, and 36e-/px total noise. The acquisition camera has a neutral density (ND) filter wheel with slots of 1\%, 10\%, 25\%, 50\% ND Vis/NIR, and a blank to accommodate high dynamic range of acquiring bright and faint stars.

\subsection{Alignment strategy}
The alignment of optics is automated in several systems. Table~\ref{Actuator_control} presents the alignment strategy, motor or actuator selection, and controller driving the motor or actuator.

\begin{table}[ht]
\caption{Opto-mechanical alignment configuration and motor/actuator types.} 
\label{Actuator_control}
\begin{center}       
\begin{tabular}{lllll}
\hline
Optics & Motion type & Precision & Repeatability & Motor and controller  \\
\hline
ADC1 & Rotational & Yes & No & Stepper, MAXnet driver\\
ADC2 & Rotational & Yes & No & Stepper, MAXnet driver\\
K-Mirror & Rotational & Yes & Yes & Stepper, MAXnet driver \\
VPyWFS fold & Linear & Yes & Yes & Stepper, MAXnet driver\\
VPyWFS Lens & XY Linear & Yes & No & PI P-625.2CD, E-727 controller\\
PI Modulation & Tip/Tilt & Yes & Yes & PI S-331.2SH, E-727 controller\\
Dichroic 1 Periscope & Tip/Tilt &Yes & Yes & Stepper, MAXnet driver\\
Dichroic 2  Periscope & Tip/Tilt, Linear & Yes & Yes & Stepper, MAXnet driver\\
Acquisition Filter Wheel & Rotational & No & No & Stepper, MAXnet driver\\
Acquisition Fold Filter Wheel & Rotational & No & No & Stepper, MAXnet driver\\
\hline
\end{tabular}
\end{center}
\end{table}

\subsection{Instrument control software}\label{software-design}
\subsubsection{Computer hardware}
\begin{figure}[ht]
\centering
\includegraphics[width=\textwidth]{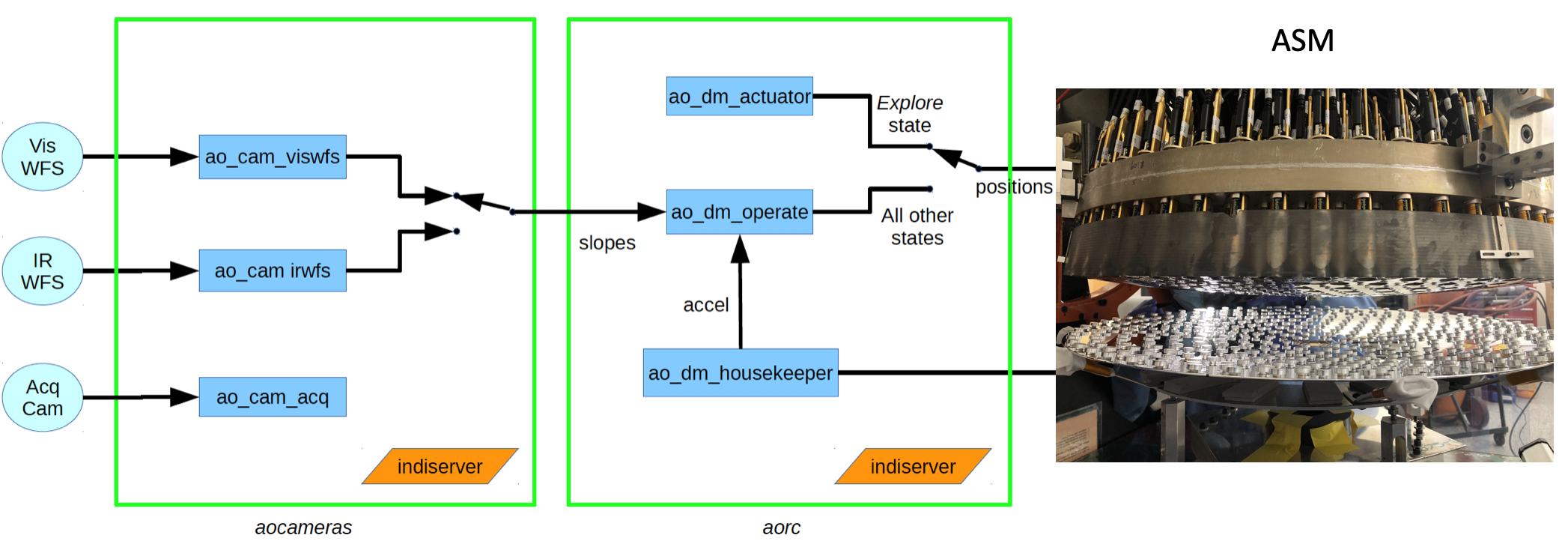}
\caption[example]{\label{computer_archi} Computer architecture of MAPS AO system. The frame grabbers of the VPyWFS-CCID75 and the IRPyWFS-SAPHIRA and the RJ45 Ethernet of the acquisition camera are installed on the \texttt{aocameras} server computer. The INDI instrument drivers, \texttt{ao\_cam\_viswfs}, \texttt{ao\_cam\_irwfs} and \texttt{ao\_cam\_acq}, which are installed on \texttt{aocameras} computer, read frames from all the cameras. The \texttt{ao\_cam\_viswfs} and \texttt{ao\_cam\_irwfs} compute slopes from the detector images and then send to \texttt{aorc} over a dedicated Ethernet socket in a ``WFS loop" operated at the WFS frame rate, 1kHz. \texttt{ao\_dm\_operate} driver works on the \texttt{aorc} computer runs a second loop ``ASM loop” independent of the WFS loop and converts the WFS slopes to the ASM commands and send to the ASM actuators to apply correction.  The \texttt{ao\_dm\_actuator} allow engineering with actuators and the \texttt{ao\_dm\_housekeeper} does house keeping and safety. 
}
\end{figure}

\begin{figure}
\centering
\includegraphics[width=\textwidth]{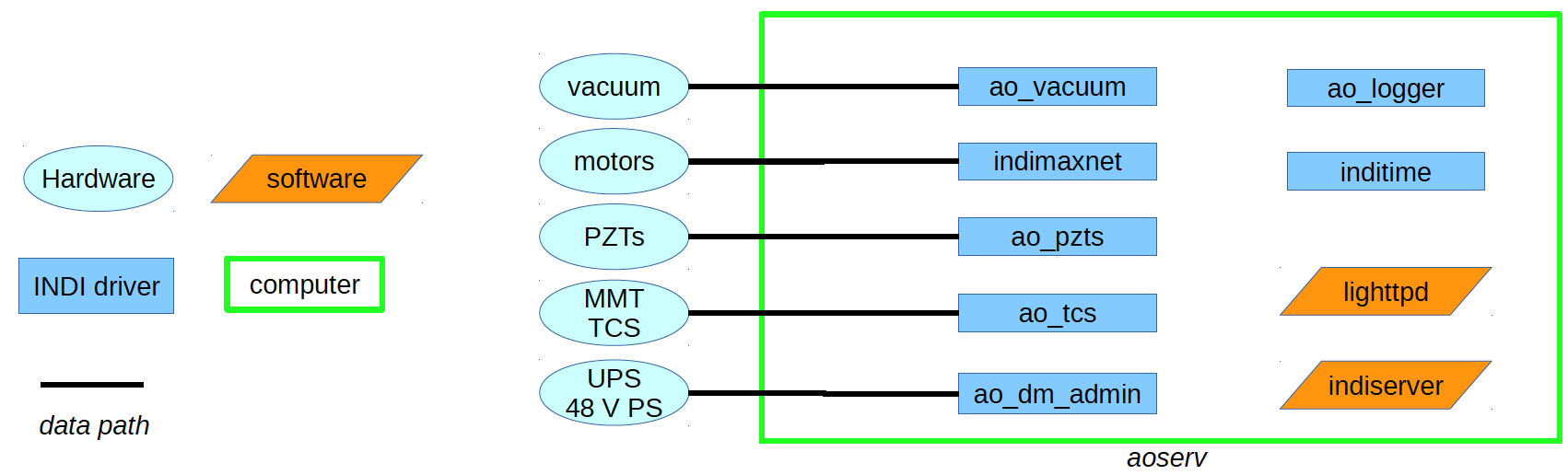}
\caption[example]{\label{aoserv_archi} The \texttt{aoserv} computer handles non-real-time Input and Output devices such as motors and actuators, web Graphical User Interfaces (GUI) and telemetry storage. For example, \texttt{lighttpd} is web socket server. The \texttt{indimaxnet} is an INDI driver for controlling motors. The \texttt{ao\_dm\_adm} is an INDI driver allows to play with ASM administrative states.}
\end{figure}

The MAPS computers are COTS products. Three computers are used for the MAPS AO control (See Figures~\ref{computer_archi} and \ref{aoserv_archi}): the AO cameras (\texttt{aocameras}), the AO Reconstruction Computer (\texttt{aorc}) and the AO server computer (\texttt{aoserv}). These computers are identical, i.e., use the same motherboard, processor, and RAM. These computers each are Intel Xeon with four cores and have 32~GB RAM.  The advantage of using the COTS parts is that they allow managing aged or non-working parts with replacements likely available for some time in the future. They also allow easy installation of software upgrades on the replacement hardware.

The software architecture is based on INDI - Instrument Neutral Distributed Interface (Elwood 2007). An example application of this architecture is the Large Binocular Telescope Interferometer instrument\cite{Pinna2016SPIE.9909E..3VP, Ertel2020}. In this architecture, servers communicate with the hardware, publish relevant information, and appropriate web-based interface clients to subscribe to the information. We do not have a central database system. All the server software is implemented in C/C++, and web-based Graphical User Interfaces (GUIs) are implemented in Java and HTML5.

Figure~\ref{computer_archi} presents the architecture of the low-latency real-time WFS loop and DM loops. These processes use high priority, locked-down memory, and fixed processor affinity for the low time-delay control loop performance. The \texttt{aocameras} is the computer where all the frame grabbers of the CCID75 and SAPHIRA cameras are installed. The INDI drivers, \texttt{ao\_cam\_viswfs} and \texttt{ao\_cam\_irwfs}, which are installed on the \texttt{aocameras} read frames from the CCID75 and SAPHIRA detectors. These INDI drivers also compute slopes and send them to \texttt{aorc} over a dedicated Ethernet socket in a ``WFS loop", operated at the WFS frame rate, 1kHz. The \texttt{aorc} server computer runs a second loop, ``ASM loop", independent of the WFS loop and converts the WFS slopes to the ASM commands. These commands are sent to the ASM actuators to apply corrections. 

The non-real-time input and output (I/O) devices such as alignment motors and actuators (See Table~\ref{Actuator_control}), web GUIs, and telemetry storage are implemented on the \texttt{aoserv} computer.

\begin{figure}[ht]
\centering
\includegraphics[width=\textwidth]{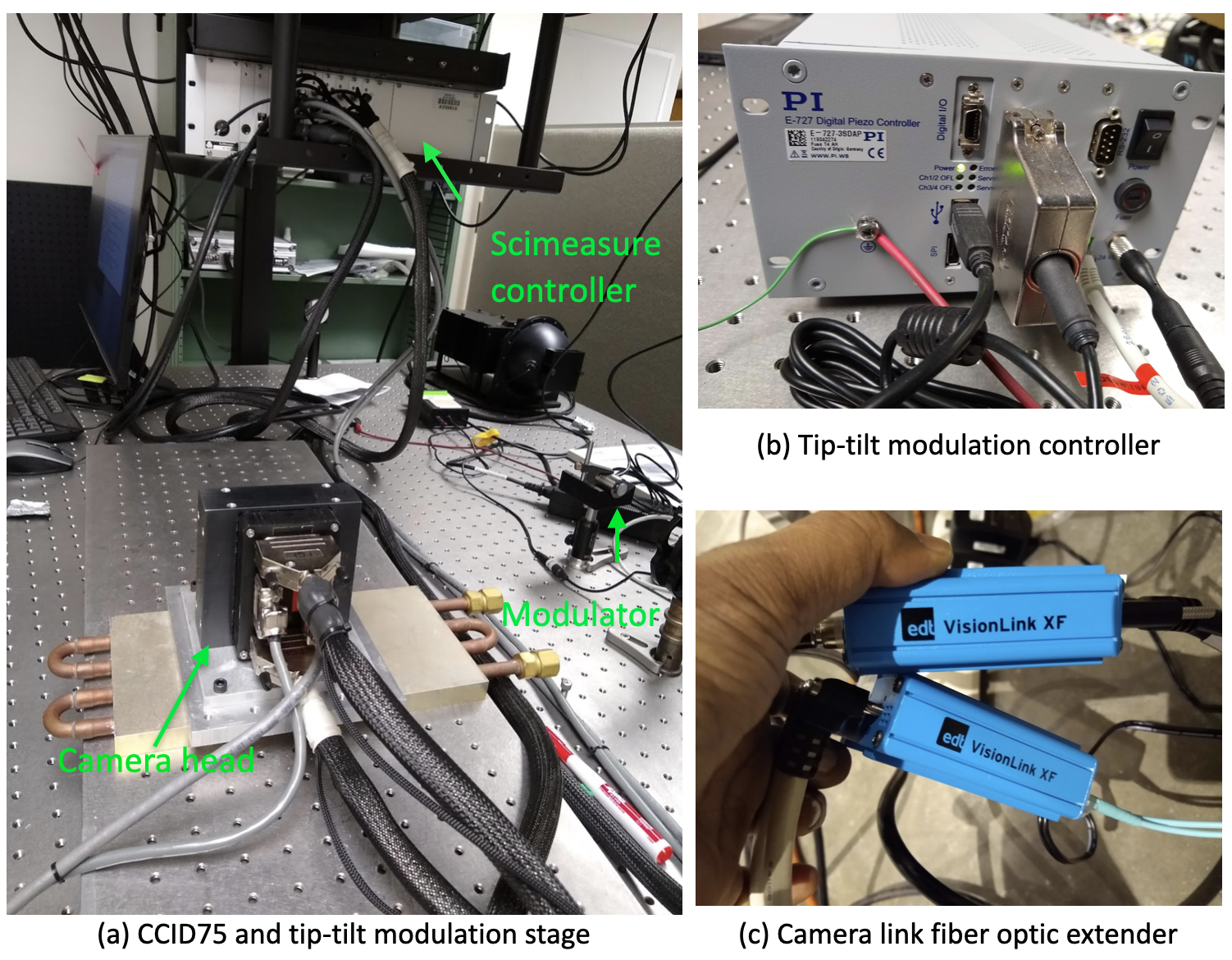}
\caption[example]{\label{lab_setup} (a) A laboratory setup to characterize the CCID75 and tip-tilt modulation mirror. (b) A  E-727 PI controller is used to control the tip-tilt modulation mirror.  (c) The camera link cables are extended several meters using a noiseless VisionLink camera-fiber-link extender system using LC/LC duplex multimode fiber optic patch cables between the CCID75 camera (will be inside the telescope) and the data acquisition computer (will be placed in the computer room) for the data acquisition. }
\end{figure}

\textbf{CCID75} is controlled by SciMeasure's Little Joe readout electronics and controller. The readout frames from the SciMeasure controller are grabbed using a low-latency optimized software from EDT  Vision Link F4 frame grabber. The EDT frame grabber software is installed on a CentOS Linux 7.8 with kernel running 3.10.0. The pixels read from the sensor are deinterlaced to reconstruct the correct frame. The software works in Linux, CentOS 7, and is developed in our lab in the INDI framework in C/C++. The camera link cables are extended several meters using a noiseless VisionLink camera-fiber-link extender system using LC/LC duplex multimode fiber optic patch cables between the CCID75 camera (will be inside the telescope) and the data acquisition computer (will be placed in the computer control room) for the data acquisition. (see Figure~\ref{lab_setup}).  

Two threads run in the \texttt{ao\_cam\_viswfs} or \texttt{ao\_cam\_irwfs} process. One thread runs the ``WFS loop" -- reads frames from the camera frame grabber and computes slopes from the detector image and sends the computed slopes  to the \texttt{aorc} computer. Second loop runs a INDI communication thread responding to user requests such as changing camera configuration, change frame rate and gain.

\textbf{Acquisition camera} is a GigE industrial camera, and we use Pylon SDK from Basler for the data acquisition. The camera control and grabbing of frames are implemented over a TCP protocol using an RJ45 Ethernet cable. 

\textbf{Fast steering tip-tilt modulator} software is implemented in Linux using E-727 library and works over a TCP protocol using an RJ45 Ethernet cable. 

\textbf{Telemetry} of all the systems and all the time are recorded, including all wavefront sensors images, slopes, and reconstruction matrix.  We plan to use this information to improve AO performance -- for gain estimation\cite{canales2000gain} and predictive control\cite{males2018ground}, the experiments in post-processing to correlate the telemetry data with science data.

\begin{figure}
\centering
\includegraphics[width=0.6\textwidth]{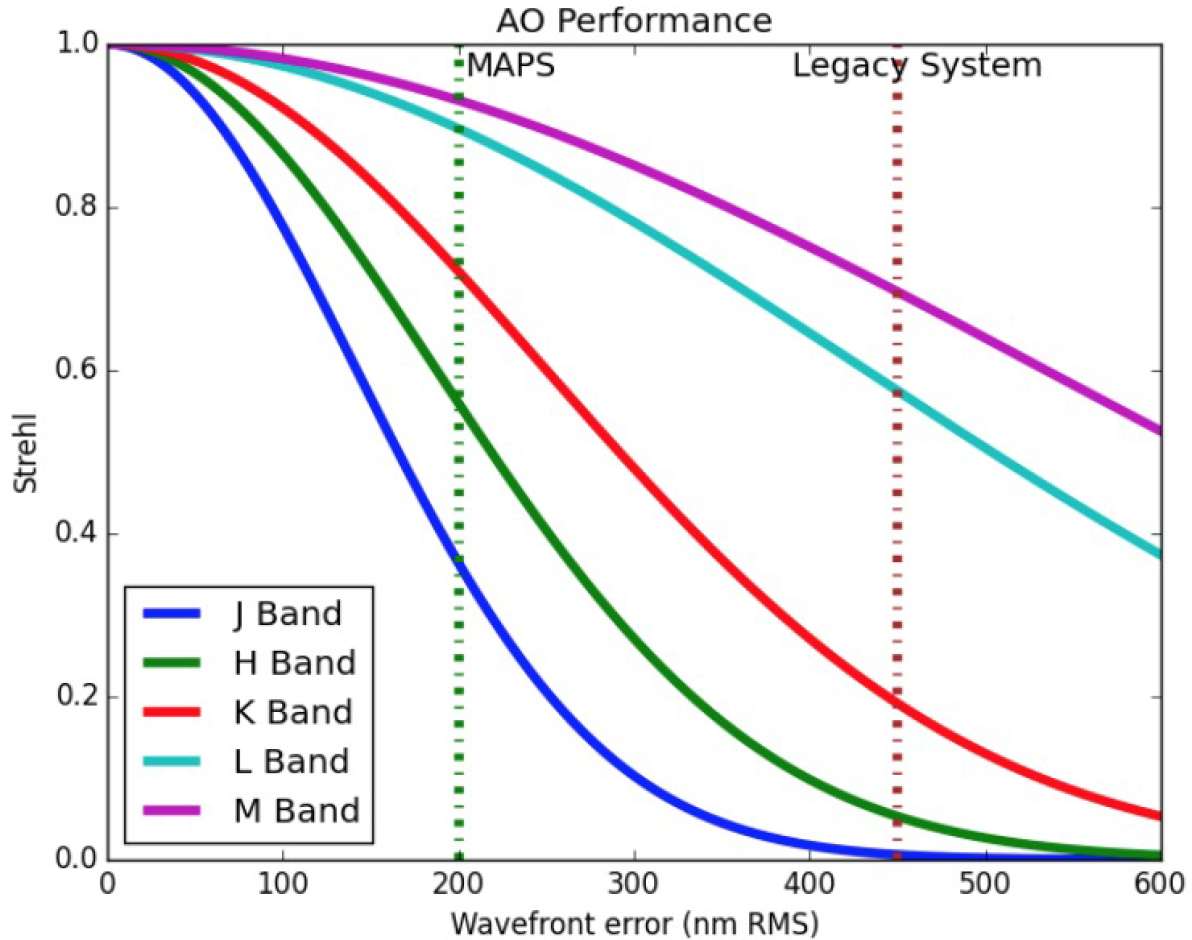}
\caption[example]{\label{Expected_performance} Expected performance of MAPS compared to legacy MMT AO system. Reproduced from Hinz et al. 2018\cite{Hinz2018}. The performance of legacy MMT AO was limited with residual error of 450~nm due to a low number 55 of corrected modes and a slow temporal response 2.7 ms delay. The new design will correct 220 modes with a faster temporal response of 1~ms delay. These efforts will push the residual wavefront error down to $<$200~nm, i.e., $>70\%$ and $60\%$ Strehl ratios at K and H-bands. }
\end{figure}

\subsection{Expected performance of MAPS}
The legacy MMT adaptive optics system was optimized for higher wavelengths with 75\% at M-band and delivered poor Strehl ratios for lower wavelengths, 20\% at H-band. The performance was limited due to a low number 55 of corrected modes and a slow temporal response 2.7 ms delay. For the median seeing at the MMT of $0.8^\prime\prime$, the residual wavefront error is approximately 450~nm RMS, i.e., Strehl ratio of  75\% at M-band and $<20$\% at H-band. In the new design, we aim several improvements to the AO control aiming to reduce the residual wavefront error and increase the Strehl ratio:

\begin{itemize}
\item  The legacy MMT AO system control was developed by Microgate, and the code was not open to making modifications to improve the performance. In MAPS, we have full freedom as we develop the control system for the ASM and wavefront sensor. We will deploy an optimal PID controller\cite{Powell2010SPIE.7736E..36P}, higher spatial mode correction, and vibration reduction using the accelerometer feed-forward to reduce residual wavefront error further and increase the Strehl ratio. This new system will correct 220 modes with a faster temporal response of 1~ms delay.  MAPS is expected to deliver two times better performance -- residual wavefront error of approximately 200~nm RMS.  These efforts will push the residual wavefront error down to $<$200~nm, i.e., $>70\%$ and $60\%$ Strehl ratios at K and H-bands (See Figure~\ref{Expected_performance}). 

\item The software is carefully designed by leveraging recent advances in the field to make it reliable and efficient to maximize scientific output. We use very advanced algorithms for the slope estimation and accurate alignment of camera triplet lens -- from the experience of LBTI\cite{Pinna2016SPIE.9909E..3VP} and MagAO-X\cite{Males2018SPIE10703E..09M}.  Simplified operational interfaces are implemented leveraging the INDI protocol aiming at improving the observational efficiency.

\item Since the deformable mirror being the secondary mirror of the telescope, there is no artificial calibration source allowing to measure an interaction matrix as done in classical AO systems. We will use a pseudo-synthetic interaction matrix reconstructor tweaked from synthetic and on-sky interaction matrices \cite{Oberti2006SPIE.6272E..20O}(lead: Masen Lamb). 
\end{itemize}

Figure~\ref{Expected_performance} shows the MAPS estimated Strehl ratio as a function of residual wavefront error and wavelength.

\section{Characterization and results}\label{Characterization}
To characterize the performance of the CCID75 camera, PI fast steering tip-tilt modulation mirror and its E-727 controller, and the Basler acquisition camera are installed in the MAPS laboratory (see Figure~\ref{lab_setup}) at the Steward Observatory, University of Arizona. 

\subsection{CCID75 wavefront sensor camera}
\begin{figure}[ht]
\centering
\includegraphics[width=0.48\textwidth]{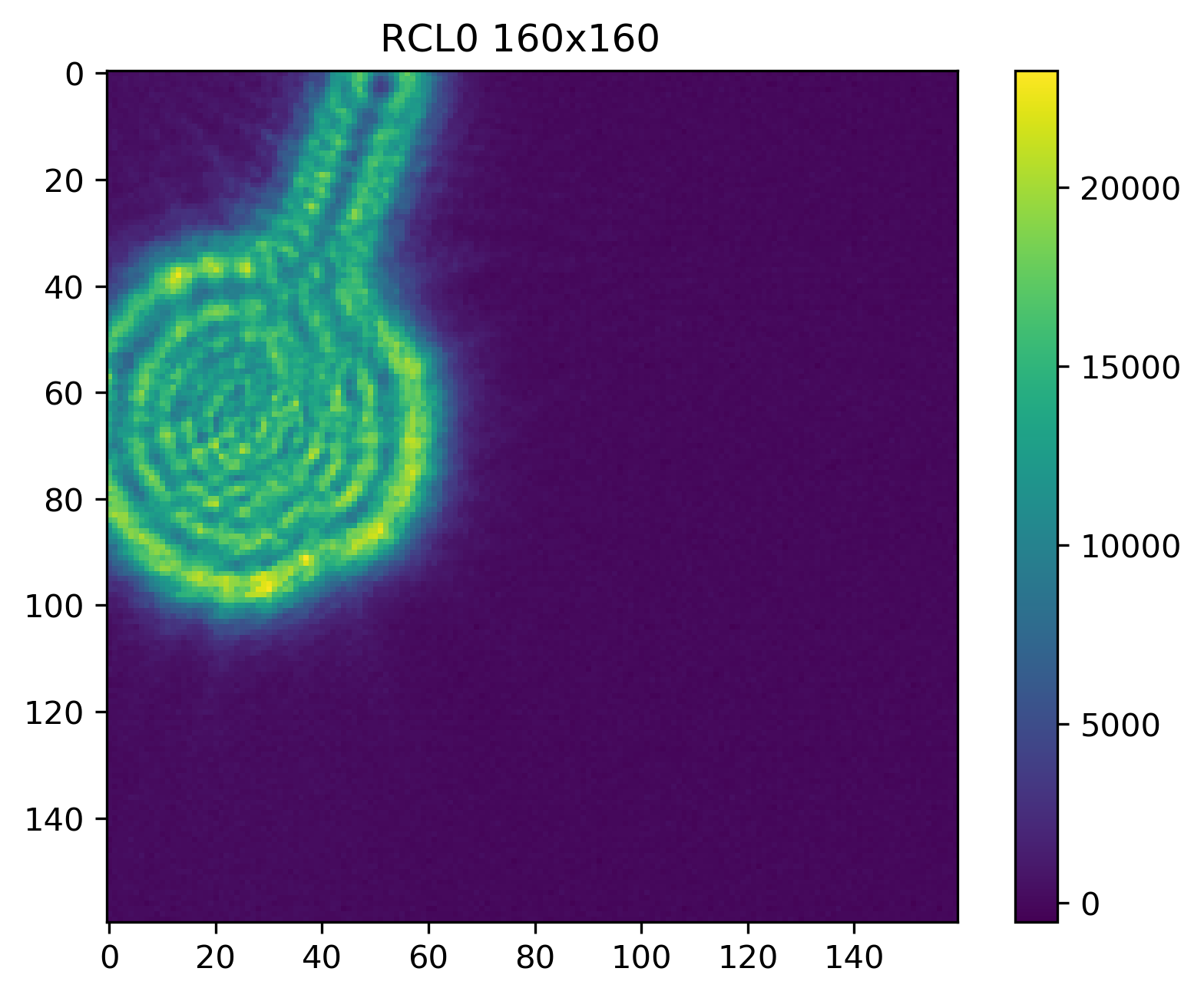}
\includegraphics[width=0.48\textwidth]{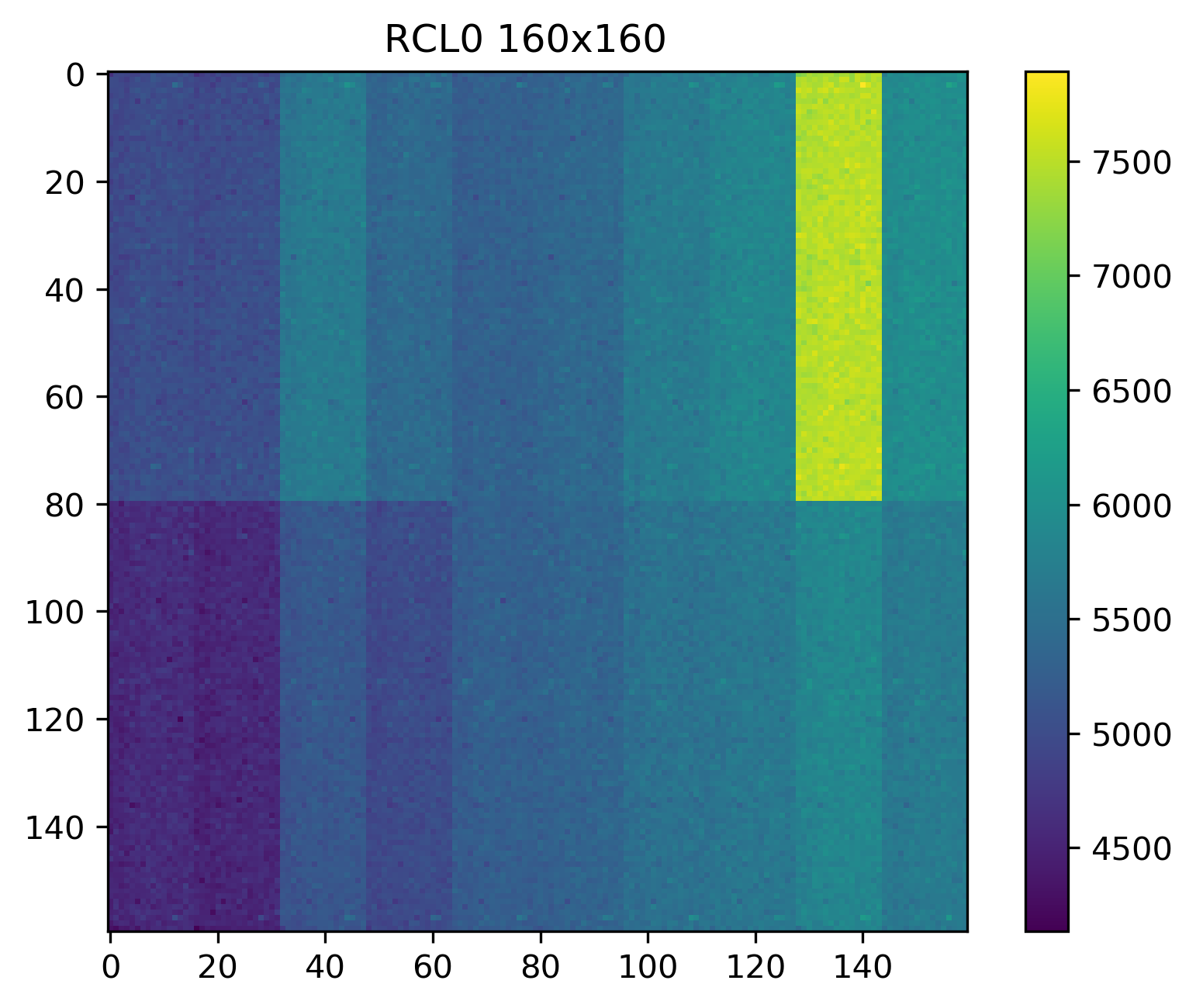}
\caption[example]{\label{ccid75_image}Scene and bias images. The bias image show 20 readout outputs with each output associated with an array of 16 (Horizontal) $\times 80$ (Vertical) imaging pixels. The bias voltages are adjusted to make them even. The wavefront sensor images will be subtracted with median bias. The bias can be dependent on the temperature fluctuations, we plan to a bias image for each target slew. }
\end{figure}

The CCID75 is characterized for its readout noise,  system gain, and linearity from the Photon Transfer Curve (PTC)\cite{li2016assessing}. Figures~\ref{ccid75_image} show the images of CCID75 that were recorded using a scene in front of the camera and bias by closing the shutter. The bias images were stable when tested by power cycling the camera and the SciMeasure controller. However, the expected temperature-dependent controller bias characterization has not been done yet. 

\begin{figure}[ht]
\centering
\includegraphics[width=0.48\textwidth]{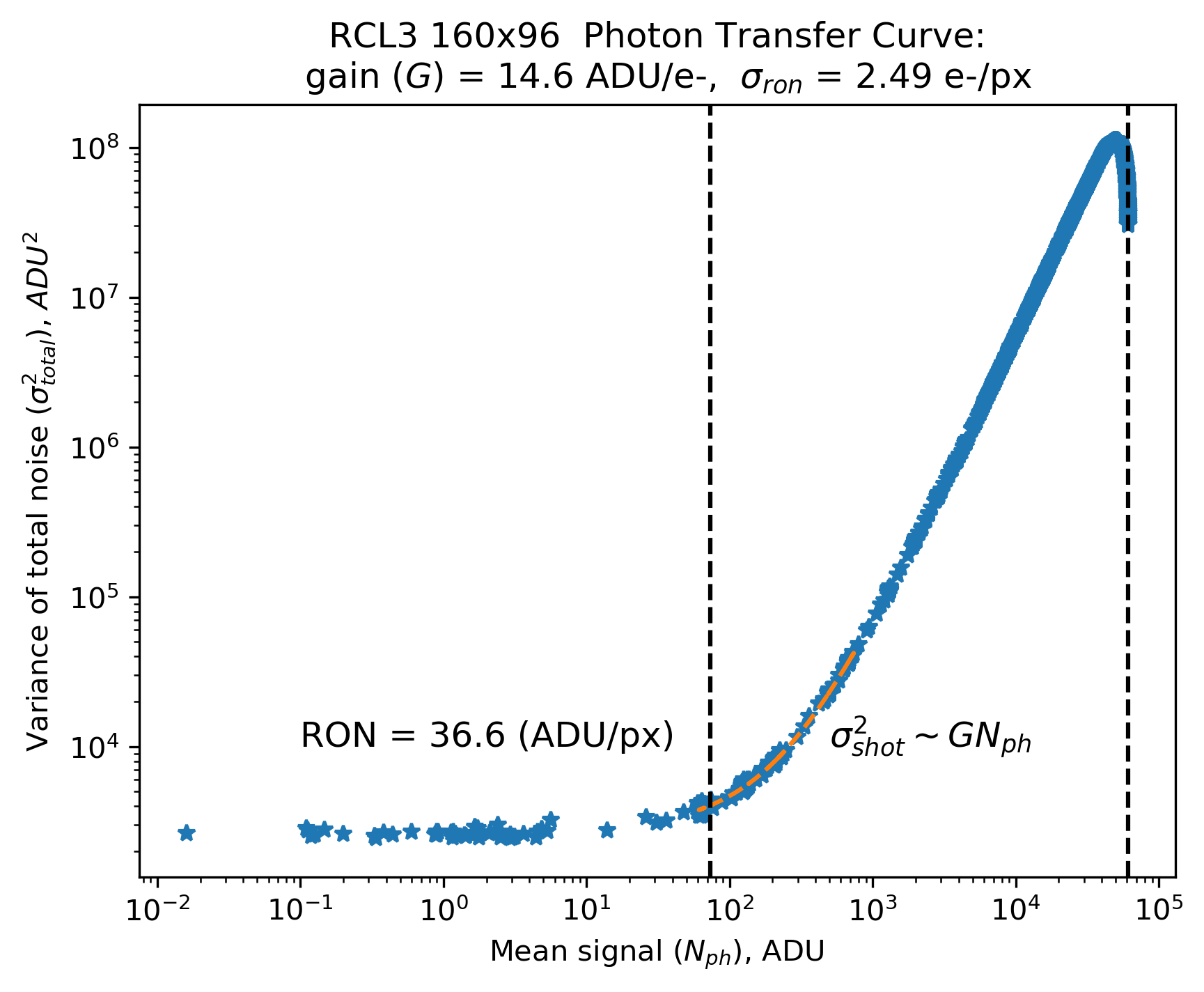}
\includegraphics[width=0.48\textwidth]{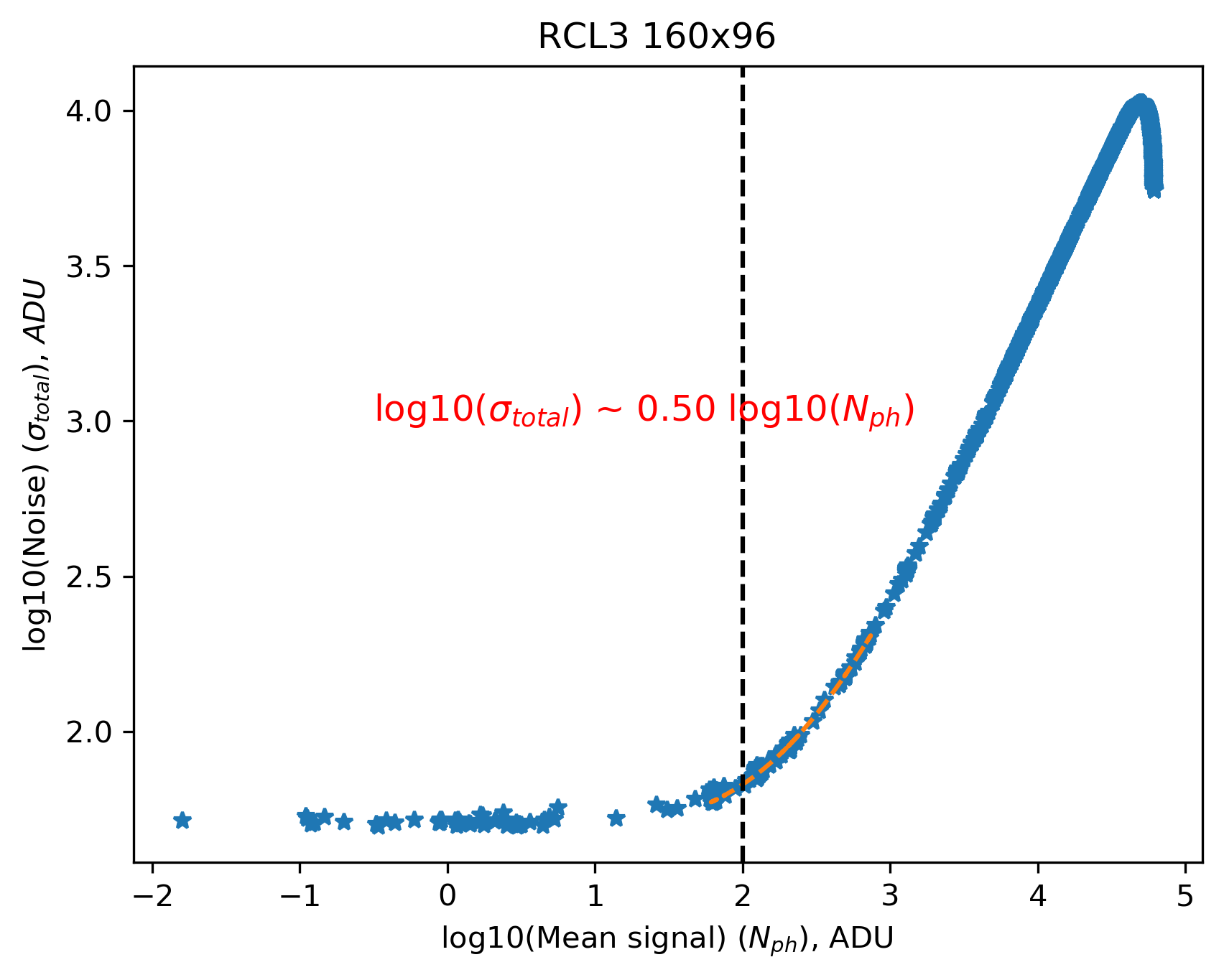}
\caption[example]{\label{plot_var_vs_signal}   (Left) PTC curve is a log‑log plot of noise versus signal. In this method, the gain and readout noise of the CCID75  are calculated by using a linear range of plot and slope where photon shot noise dominates. PTC shows various noise regions (readout noise, shot-noise, and saturated region). The RMS value of shot noise is equal to the square root of the photon incident's mean number on a given pixel. Thus, the shot noise profile becomes a straight line with a slope of 0.5 on the log-log curve. Keep in mind that shot noise is inherent in the nature of light itself and has nothing to do with the camera design. (Right)  The camera gain, ADU/e-, is obtained by fitting a line to the variance vs. mean signal curve in the shot (photon) noise limited region and measuring the slope of that line.}
\end{figure}

\begin{itemize}
\item For the PTC measurement, the CCID75 is exposed to a precisely controlled laser source of light and illuminated uniformly flat on the detector. Once the noise at each illumination level is calculated, a PTC is generated by plotting the camera's output RMS noise vs mean signal level on a log-log curve (see Figure~\ref{plot_var_vs_signal}).

\item Read noise is directly available from this PTC by recording the noise level at zero illumination.

\item The camera gain is obtained by fitting a line to the variance vs mean signal curve in the shot noise limited region and by measuring the slope of that line. The gain of camera systems is typically expressed in ADU/e-, that is, the number of analog to digital units per signal electron (see Figure~\ref{plot_var_vs_signal}). 
    
\item The readout noise in ADU/px is divided by the system's gain to yield the noise in e-/px. 
\end{itemize}

\begin{figure}[ht]
\centering
\includegraphics[width=0.48\textwidth]{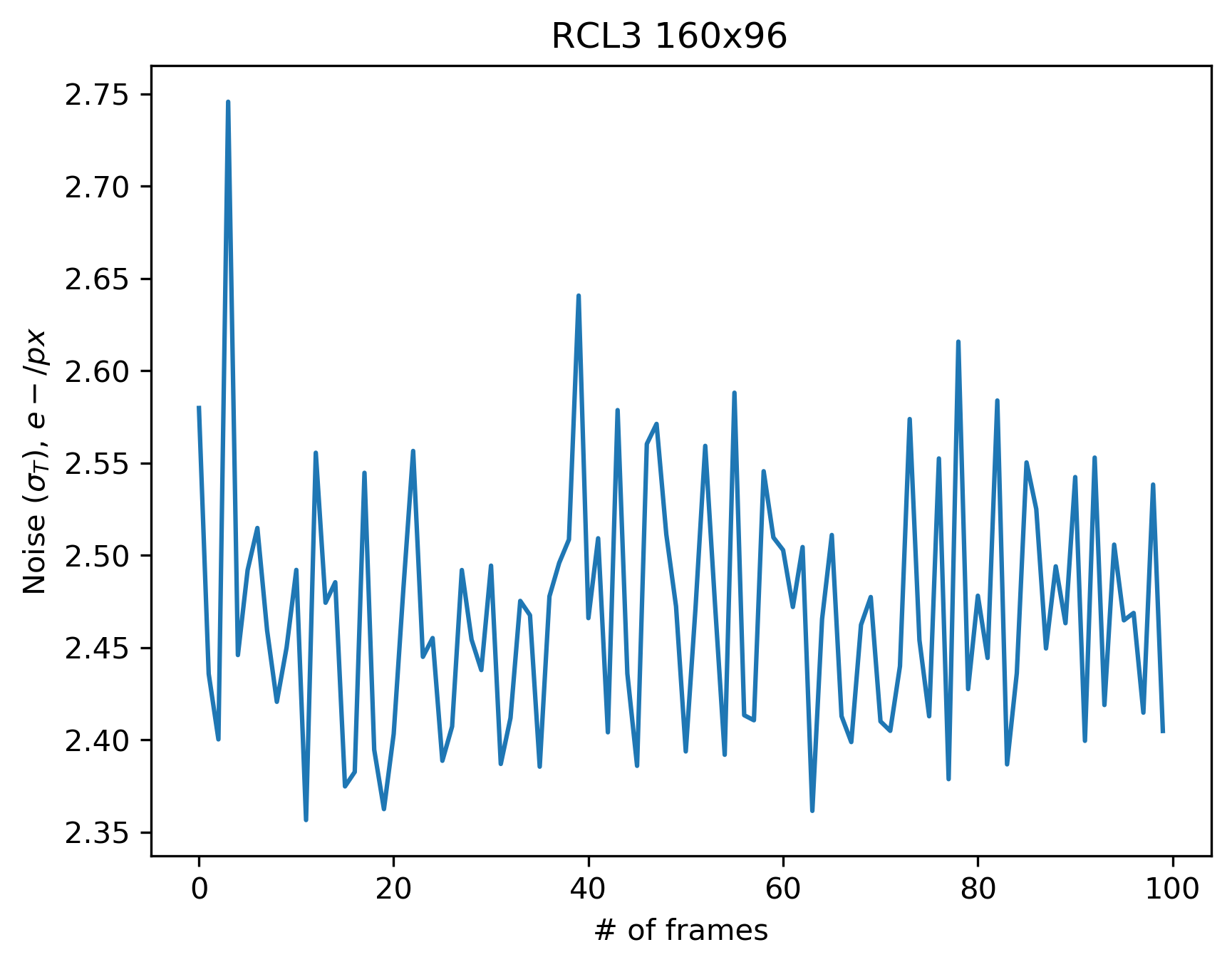}
\includegraphics[width=0.48\textwidth]{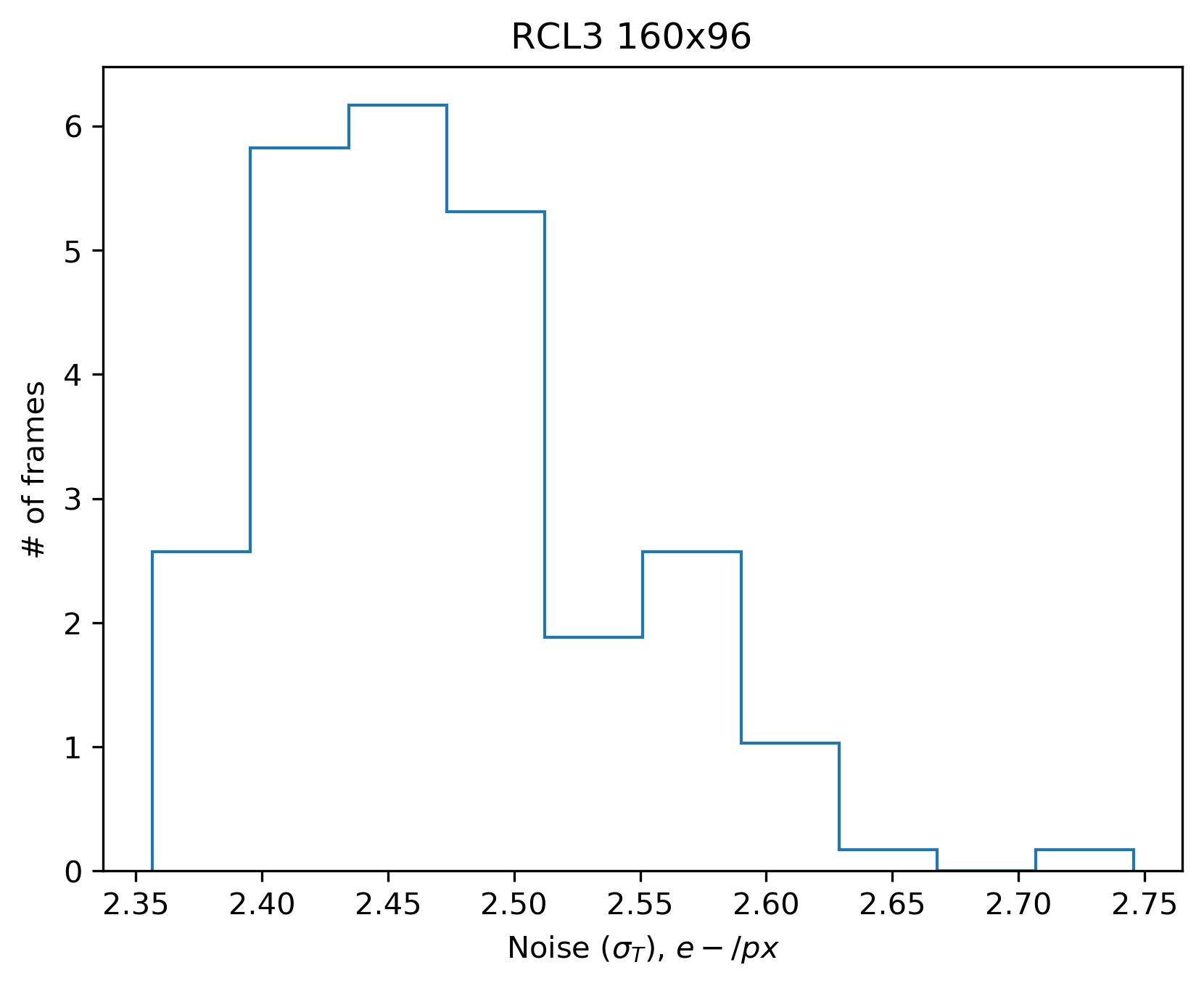}
\caption[example]{\label{RON_measured_from_STD}Readout noise, $\sigma_{\rm R}$ for various frames.}
\end{figure}

Table~\ref{Table:1} presents various \texttt{RCL} programs of CCID75 and corresponding ROI, system gain, readout noise, dynamic range, and frame rates. The measured readout noise is more or less matches with the expected readout noise numbers from the company\cite{Schuette2014}.   There is a small difference in the readout noise measurement frame to frame, and that is shown in Figure~\ref{RON_measured_from_STD}. Furthermore, we measured quantum efficiency by using a calibrated black-body integrated sphere, and the measured quantum efficiency matches with Schuette et al. 2014\cite{Schuette2014} with $\pm5\%$ deviation. 

We need $80\times80$ pixels ROI for no binning mode, and we will use \texttt{RCL3} as it offers the best results required for our specifications: $160\times96$ pixels ROI, 2.5e-/px readout noise at 2kHz frame rate. We trim the unused pixels in the VPyWFS software and make final $80\times80$ ROI. For $2\times2$ binning, we have optional readout $80 \times 80$ ROI, e.g., \texttt{RCL12} with readout noise 3e-/px. For $4\times4$ binning, we have optional  $40\times40$ ROI, e.g., \texttt{RCL14} with readout noise 2e-/px.

\begin{table}
\caption[example]{CCID75 characterization results. \texttt{RCL3} 
will be selected for no-binning readout mode case.}             
\label{Table:1}      
\centering                          
\begin{tabular}{c c c c c c c c}
\hline
Program & Readout  & Gain  & $\sigma^{\rm mes}_{\rm R}$  &  $\sigma^{\rm exp}_{\rm R}$ & Dynamic range  &  Frame rate   & Pixel Rate  \\
 & size  &  ADU/e- &  e-/px &  e-/px &  ADU/px &   Hz  & MHz \\
\hline
\texttt{RCL} 0 & $160\times160$  & 13.8 & 3.69 & 4.4 & 60455 & 2151& 4.54\\
\texttt{RCL} 1 & $160\times128$  & 13.6 & 4.04 & 3.9 & 62262 & 2287 & 3.85 \\
\texttt{RCL} 2 & $160\times160$  & 15.8 & 2.86 & 2.7 & 62675 & 1441 & 2.78 \\
\texttt{RCL} 3 & $160\times96$   & 14.6 & 2.49 & 2.7 & 62395 & 2303 & 2.78 \\
\texttt{RCL} 4 & $160\times128$  & 12.3 & 3.60 & 2.1 & 63261 & 1014 & 1.47\\
\texttt{RCL} 5 & $160\times96$   & 15.8 & 2.54 & 2.1 & 62948 & 1332 & 1.47\\
\texttt{RCL} 6 & $160\times96$   & 17.6 & 2.94 & 2.0 & 62932 & 931  & 1.01 \\
\texttt{RCL} 7 & $160\times128$  & 11.0 & 5.63 & 5.9 & 42685 & 3083 & 5.56\\
\texttt{RCL} 8 & $160\times160$  & 8.23 & 3.73 & 4.4 & 35103 & 2151 & 4.54\\
\texttt{RCL} 16 & $160\times160$ & 0.94 & 7.45 & 4.4 & 4999  & 2151 & 4.54\\
\texttt{RCL} 24 & $160\times160$ & 0.73 & 6.30 & 4.4 & 4815  & 2151 & 4.54\\
\hline
\end{tabular}
\vskip12pt
\end{table}

The pyramid wavefront sensor images need to be synchronized with the modulator motions -- the camera integration period should be an integer multiple of the modulation period. We are currently working on the synchronization of the CCID75 wavefront sensor camera images with the PI E-727 modulator tip-tilt motions.  The SciMeasure controller of the CCID75 camera generates an output signal for each frame, and this external trigger is used in the PI E-727 controller to move the tip-tilt modulator mirror.

\subsection{Tip-tilt modulation mirror}
\begin{figure}[ht]
\centering
\includegraphics[width=\textwidth]{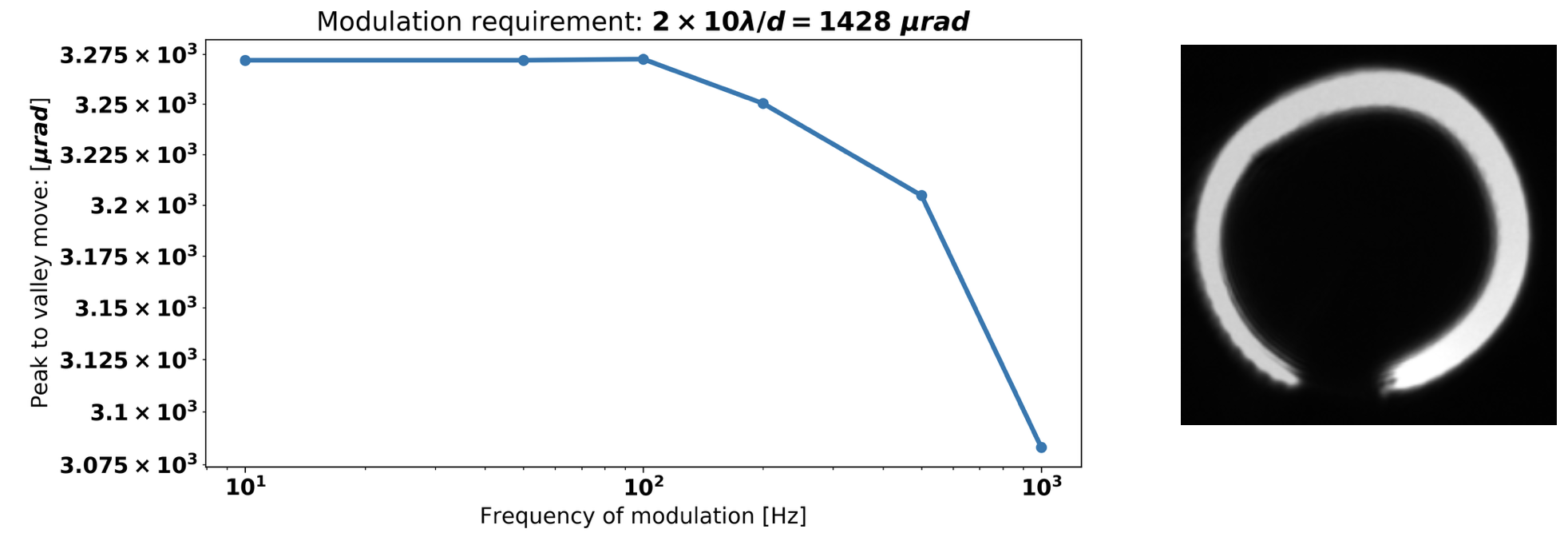}
\caption[example]{\label{modulation_frequncy_response} (Left) Bode plot -- frequency response of modulation system showing modulation amplitude for various modulation frequencies. The system is capable of delivering modulation two times the requirement. (Right) $3000\mu$rad amplitude and 1000~Hz modulation  circle recorded on the CCID75 camera using a laser beam reflected on the modulation system.}
\end{figure}

The fast steering tip-tilt modulation system is characterized for its frequency response -- checking the peak-to-valley modulation amplitude at several operational frequencies. We verified the modulation amplitude physically by installing a laser source, reflected the laser beam on the tip-tilt modulation system, and recorded a circular modulation on the CCID75. Our requirement of $2\times 10\lambda/D$ modulation is equivalent to peak-to-valley of $1428\mu$rad. The system is capable of delivering more than two times our required modulation amplitude (i.e., $3000\mu$rad). Figure~\ref{modulation_frequncy_response} shows the frequency response and recorded circular modulation. 

\begin{figure}[ht]
\centering
\includegraphics[width=\textwidth]{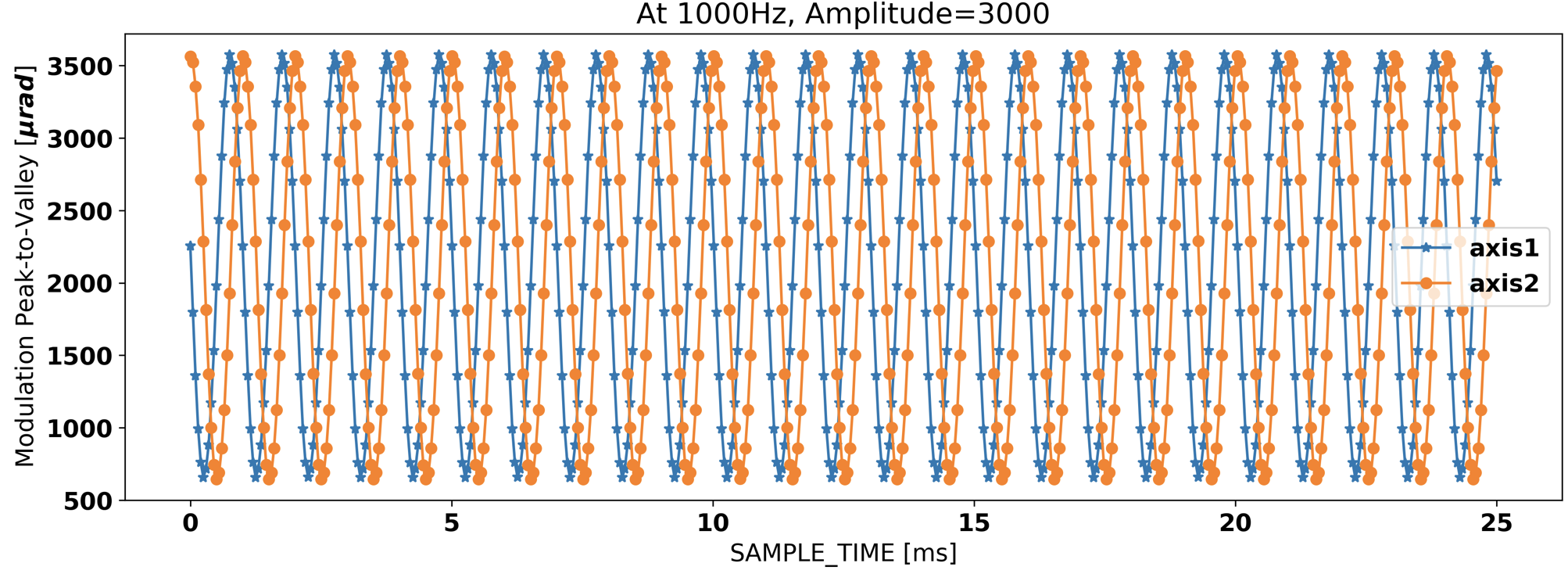}
\caption[example]{\label{modulation} A circular modulation with $3000\mu$rad at 1000~Hz is created with the tip-tilt modulation stage by applying 90-degree phase-shift between two axes moving with sine wave.}
\end{figure}

The circular modulation is created by applying sine-wave-generators to the two tip-tilt axes and a 90-degree phase-shift between them (see Figure~\ref{modulation}). The factory-built sine-wave-generator is not sampled very well, and that caused limitations in: (i) making a perfectly circular modulation; (ii) making modulation for all frequencies and amplitudes we wanted.  To fix this, we generate sin-wave data in our software and write that to the firmware of the E-727 controller.

\section{Summary and Future work}
We presented an overall design of the MAPS visible pyramid wavefront sensor.    The final design of the wavefront sensor is post-PDR. The optics is under procurement, and they are expected to deliver by the end of 2020. Once the optics arrive, we will machine mechanical parts. Then we will start the integration of optics on the VPyWFS optical bench at the Steward Observatory and test the optics and verify the software, which is already working with simulated wavefront sensor images. 

An AO system's success depends on lots of testing the cameras and modulation devices and with the software, and we have carried out all possible tests. We have already installed and characterized the critical components of the VPyWFS system: the Basler acquisition camera, PI fast steering tip-tilt modulator, and CCID75 wavefront sensor camera in the lab. All these systems are within the specifications.

The MAPS-ASM development is finished and almost ready for usage with the wavefront sensor. On sky-tests and commissioning on the MMT is planned for summer 2021. We will use a pseudo-synthetic interaction matrix reconstructor that is tweaked from simulations and an on-sky interaction matrix. MAPS exploits the low thermal emissivity of ASM, ARIES, and MMTPol and delivers unique astronomical capabilities. 

\acknowledgments 
The MAPS project is primarily funded through the NSF Mid-Scale Innovations Program, programs AST-1636647 and AST-1836008.

\bibliography{main} 

\begin{thebibliography}{10}

\bibitem{Morzinski2020}
Morzinski, K.~M., Montoya, M., Fellows, C., Durney, O., Ford, J., West, G.,
  Gardner, A., Vaz, A., Anugu, N., Mailhot, E., Carlson, J., Harrison, L.,
  Gacon, F., Downey, E., Jones, T., Hinz, P., Patience, J., Sivanandam, S.,
  Chen, S., Lamb, M., Butko, A., Liu, S., Hardy, T., and Jannuzi, B.,
  ``{Development and status of MAPS, the MMT AO exoPlanet characterization
  system},'' in [{\em This Proceedings}{\nolinebreak\hspace{0.1em}]},  {\em
  Society of Photo-Optical Instrumentation Engineers (SPIE) Conference Series}
  {\bf 11448},  11448--61 (2020).

\bibitem{Williams2018SPIE10700E..2TW}
{Williams}, G.~G., ``{The MMT Observatory: entering a new era of scientific
  discovery},'' in [{\em Ground-based and Airborne Telescopes
  VII}{\nolinebreak\hspace{0.1em}]},  {Marshall}, H.~K. and {Spyromilio}, J.,
  eds., {\em Society of Photo-Optical Instrumentation Engineers (SPIE)
  Conference Series} {\bf 10700},  107002T (July 2018).

\bibitem{Lloyd-Hart2000SPIE.4007..167L}
{Lloyd-Hart}, M., {Wildi}, F.~P., {Martin}, B., {McGuire}, P.~C., {Kenworthy},
  M.~A., {Johnson}, R.~L., {Fitz-Patrick}, B.~C., {Angeli}, G.~Z., {Miller},
  S.~M., and {Angel}, J. R.~P., ``{Adaptive optics for the 6.5-m MMT},'' in
  [{\em Adaptive Optical Systems Technology}{\nolinebreak\hspace{0.1em}]},
  {Wizinowich}, P.~L., ed., {\em Society of Photo-Optical Instrumentation
  Engineers (SPIE) Conference Series} {\bf 4007},  167--174 (July 2000).

\bibitem{Brusa2003SPIE.4839..691B}
{Brusa}, G., {Riccardi}, A., {Salinari}, P., {Wildi}, F.~P., {Lloyd-Hart}, M.,
  {Martin}, H.~M., {Allen}, R., {Fisher}, D., {Miller}, D.~L., {Biasi}, R.,
  {Gallieni}, D., and {Zocchi}, F., ``{MMT adaptive secondary: performance
  evaluation and field testing},'' in [{\em Adaptive Optical System
  Technologies II}{\nolinebreak\hspace{0.1em}]},  {Wizinowich}, P.~L. and
  {Bonaccini}, D., eds., {\em Society of Photo-Optical Instrumentation
  Engineers (SPIE) Conference Series} {\bf 4839},  691--702 (Feb. 2003).

\bibitem{Hinz2018}
{Hinz}, P.~M., {Downey}, E., {Montoya}, O.~M., {Ford}, J., {Powell}, K., and
  {Hill}, R., ``{Developing new adaptive secondary electronics for the MAPS
  project},'' in [{\em Adaptive Optics Systems
  VI}{\nolinebreak\hspace{0.1em}]},  {\em Society of Photo-Optical
  Instrumentation Engineers (SPIE) Conference Series} {\bf 10703},  1070369
  (July 2018).

\bibitem{McCarthy1998SPIE.3354..750M}
{McCarthy}, D.~W., {Burge}, J.~H., {Angel}, J. R.~P., {Ge}, J., {Sarlot},
  R.~J., {Fitz-Patrick}, B.~C., and {Hinz}, J.~L., ``{ARIES: Arizona infrared
  imager and echelle spectrograph},'' in [{\em Infrared Astronomical
  Instrumentation}{\nolinebreak\hspace{0.1em}]},  {Fowler}, A.~M., ed., {\em
  Society of Photo-Optical Instrumentation Engineers (SPIE) Conference Series}
  {\bf 3354},  750--754 (Aug. 1998).

\bibitem{Jones2007AAS...211.1124J}
{Jones}, T.~J. and {Packham}, C.~C., ``{MMTPol, A High Precision Imaging
  Polarimeter for the MMT},'' in [{\em American Astronomical Society Meeting
  Abstracts}{\nolinebreak\hspace{0.1em}]},  {\em American Astronomical Society
  Meeting Abstracts} {\bf 211},  11.24 (Dec. 2007).

\bibitem{Lloyd-Hart2000PASP..112..264L}
{Lloyd-Hart}, M., ``{Thermal Performance Enhancement of Adaptive Optics by Use
  of a Deformable Secondary Mirror},'' {\em Publ. Astron. Soc. Pac}~{\bf 112},
  264--272 (Feb. 2000).

\bibitem{Birkby2017AJ....153..138B}
{Birkby}, J.~L., {de Kok}, R.~J., {Brogi}, M., {Schwarz}, H., and {Snellen},
  I.~A.~G., ``{Discovery of Water at High Spectral Resolution in the Atmosphere
  of 51 Peg b},'' {\em Astrophysical Journal}~{\bf 153},  138 (Mar. 2017).

\bibitem{Liu2018}
{Liu}, S., {Sivanandam}, S., {Chen}, S., {Lamb}, M., {Butko}, A., {Veran},
  J.-P., {Hinz}, P., {Mieda}, E., {Hardy}, T., {Lardiere}, O., and {Shore}, E.,
  ``{Upgrading the MMT AO system with a near-infrared pyramid wavefront
  sensor},'' in [{\em Adaptive Optics Systems VI}{\nolinebreak\hspace{0.1em}]},
   {\em Society of Photo-Optical Instrumentation Engineers (SPIE) Conference
  Series} {\bf 10703},  107032K (July 2018).

\bibitem{Butko2018SPIE10702E..3NB}
{Butko}, A., {Sivanandam}, S., {Hardy}, T., {Liu}, S., {Chen}, S., {Veran},
  J.-P., and {Hinz}, P.~M., ``{Developing an infrared APD array camera for
  near-infrared wavefront sensing},'' in [{\em Ground-based and Airborne
  Instrumentation for Astronomy VII}{\nolinebreak\hspace{0.1em}]},  {Evans},
  C.~J., {Simard}, L., and {Takami}, H., eds., {\em Society of Photo-Optical
  Instrumentation Engineers (SPIE) Conference Series} {\bf 10702},  107023N
  (July 2018).

\bibitem{Vaz2020}
Vaz, A., Morzinski, K.~M., Montoya, M., Fellows, C., Ford, J., Gardner, A.,
  Durney, O., West, G., Harrison, L., Gacon, F., Downey, E., Carlson, J.,
  Mailhot, E., Anugu, N., Jannuzi, B., and Hinz, P., ``{Laboratory testing and
  calibration of the upgraded MMT adaptive secondary mirror},'' in [{\em This
  Proceedings}{\nolinebreak\hspace{0.1em}]},  {\em Society of Photo-Optical
  Instrumentation Engineers (SPIE) Conference Series} {\bf 11448},  11448--331
  (2020).

\bibitem{Rigaut2013aoel.confE..18R}
{Rigaut}, F. and {Van Dam}, M., ``{Simulating Astronomical Adaptive Optics
  Systems Using Yao},'' in [{\em Proceedings of the Third AO4ELT
  Conference}{\nolinebreak\hspace{0.1em}]},  {Esposito}, S. and {Fini}, L.,
  eds.,  18 (Dec. 2013).

\bibitem{Ragazzoni1996JMOp...43..289R}
{Ragazzoni}, R., ``{Pupil plane wavefront sensing with an oscillating prism},''
  {\em Journal of Modern Optics}~{\bf 43},  289--293 (Feb. 1996).

\bibitem{Ragazzoni1999A&A...350L..23R}
{Ragazzoni}, R. and {Farinato}, J., ``{Sensitivity of a pyramidic Wave Front
  sensor in closed loop Adaptive Optics},'' {\em Astronomy \&
  Astrophysics}~{\bf 350},  L23--L26 (Oct. 1999).

\bibitem{Tozzi2008SPIE.7015E..58T}
{Tozzi}, A., {Stefanini}, P., {Pinna}, E., and {Esposito}, S., ``{The double
  pyramid wavefront sensor for LBT},'' in [{\em Adaptive Optics
  Systems}{\nolinebreak\hspace{0.1em}]},  {Hubin}, N., {Max}, C.~E., and
  {Wizinowich}, P.~L., eds., {\em Society of Photo-Optical Instrumentation
  Engineers (SPIE) Conference Series} {\bf 7015},  701558 (July 2008).

\bibitem{Schuette2014}
{Schuette}, D.~R., {Reich}, R.~K., {Prigozhin}, I., {Burke}, B.~E., and
  {Johnson}, R., ``{New CCD imagers for adaptive optics wavefront sensors},''
  in [{\em Adaptive Optics Systems IV}{\nolinebreak\hspace{0.1em}]},  {\em
  Society of Photo-Optical Instrumentation Engineers (SPIE) Conference Series}
  {\bf 9148},  91485O (Aug. 2014).

\bibitem{Finger2014SPIE.9148E..17F}
{Finger}, G., {Baker}, I., {Alvarez}, D., {Ives}, D., {Mehrgan}, L., {Meyer},
  M., {Stegmeier}, J., and {Weller}, H.~J., ``{SAPHIRA detector for infrared
  wavefront sensing},'' in [{\em Adaptive Optics Systems
  IV}{\nolinebreak\hspace{0.1em}]},  {Marchetti}, E., {Close}, L.~M., and
  {Vran}, J.-P., eds., {\em Society of Photo-Optical Instrumentation Engineers
  (SPIE) Conference Series} {\bf 9148},  914817 (Aug. 2014).

\bibitem{Pinna2016SPIE.9909E..3VP}
{Pinna}, E., {Esposito}, S., {Hinz}, P., {Agapito}, G., {Bonaglia}, M.,
  {Puglisi}, A., {Xompero}, M., {Riccardi}, A., {Briguglio}, R., {Arcidiacono},
  C., {Carbonaro}, L., {Fini}, L., {Montoya}, M., and {Durney}, O., ``{SOUL:
  the Single conjugated adaptive Optics Upgrade for LBT},'' in [{\em Adaptive
  Optics Systems V}{\nolinebreak\hspace{0.1em}]},  {Marchetti}, E., {Close},
  L.~M., and {V{\'e}ran}, J.-P., eds., {\em Society of Photo-Optical
  Instrumentation Engineers (SPIE) Conference Series} {\bf 9909},  99093V (July
  2016).

\bibitem{Ertel2020}
Ertel, S., Hinz, P.~M., Stone, J.~M., Montoya, O.~M., West, G.~S., Durney, O.,
  Grenz, P., Spalding, E.~A., Leisenring, J.~M., Wagner, K., Anugu, N., Power,
  J., Maier, E.~R., Defrère, D., Hoffmann, W.~F., Perera, S., Brown, S.,
  Skemer, A.~J., Mennesson, B., Kennedy, G.~M., Downey, E.~C., Hill, J.~M.,
  Pinna, E., Puglisi, A.~T., and Rossi, F., ``{Laboratory testing and
  calibration of the upgraded MMT adaptive secondary mirror},'' in [{\em
  Optical and Infrared Interferometry and Imaging
  VII}{\nolinebreak\hspace{0.1em}]},  {\em Society of Photo-Optical
  Instrumentation Engineers (SPIE) Conference Series} {\bf 11446},  11446--4
  (2020).

\bibitem{canales2000gain}
Canales, V. and Cagigal, M., ``Gain estimate for exoplanet detection with
  adaptive optics,'' {\em Astronomy and Astrophysics Supplement Series}~{\bf
  145}(3),  445--449 (2000).

\bibitem{males2018ground}
Males, J.~R. and Guyon, O., ``Ground-based adaptive optics coronagraphic
  performance under closed-loop predictive control,'' {\em Journal of
  Astronomical Telescopes, Instruments, and Systems}~{\bf 4}(1),  019001
  (2018).

\bibitem{Powell2010SPIE.7736E..36P}
{Powell}, K.~B. and {Vaitheeswaran}, V., ``{Implementation and on-sky results
  of an optimal wavefront controller for the MMT NGS adaptive optics system},''
  in [{\em Adaptive Optics Systems II}{\nolinebreak\hspace{0.1em}]},
  {Ellerbroek}, B.~L., {Hart}, M., {Hubin}, N., and {Wizinowich}, P.~L., eds.,
  {\em Society of Photo-Optical Instrumentation Engineers (SPIE) Conference
  Series} {\bf 7736},  773636 (July 2010).

\bibitem{Males2018SPIE10703E..09M}
{Males}, J.~R., {Close}, L.~M., {Miller}, K., {Schatz}, L., {Doelman}, D.,
  {Lumbres}, J., {Snik}, F., {Rodack}, A., {Knight}, J., {Van Gorkom}, K.,
  {Long}, J.~D., {Hedglen}, A., {Kautz}, M., {Jovanovic}, N., {Morzinski}, K.,
  {Guyon}, O., {Douglas}, E., {Follette}, K.~B., {Lozi}, J., {Bohlman}, C.,
  {Durney}, O., {Gasho}, V., {Hinz}, P., {Ireland}, M., {Jean}, M., {Keller},
  C., {Kenworthy}, M., {Mazin}, B., {Noenickx}, J., {Alfred}, D., {Perez}, K.,
  {Sanchez}, A., {Sauve}, C., {Weinberger}, A., and {Conrad}, A., ``{MagAO-X:
  project status and first laboratory results},'' in [{\em Adaptive Optics
  Systems VI}{\nolinebreak\hspace{0.1em}]},  {Close}, L.~M., {Schreiber}, L.,
  and {Schmidt}, D., eds., {\em Society of Photo-Optical Instrumentation
  Engineers (SPIE) Conference Series} {\bf 10703},  1070309 (July 2018).

\bibitem{Oberti2006SPIE.6272E..20O}
{Oberti}, S., {Quir{\'o}s-Pacheco}, F., {Esposito}, S., {Muradore}, R.,
  {Arsenault}, R., {Fedrigo}, E., {Kasper}, M., {Kolb}, J., {Marchetti}, E.,
  {Riccardi}, A., {Soenke}, C., and {Stroebele}, S., ``{Advances in Adaptive
  Optics II},'' in [{\em Society of Photo-Optical Instrumentation Engineers
  (SPIE) Conference Series}{\nolinebreak\hspace{0.1em}]},  {Ellerbroek}, B.~L.
  and {Bonaccini Calia}, D., eds., {\em Society of Photo-Optical
  Instrumentation Engineers (SPIE) Conference Series} {\bf 6272},  627220 (June
  2006).

\bibitem{li2016assessing}
Luchang, L., Mengting, L., Zhang, Z., and Zhen-Li, H., ``Assessing low-light
  cameras with photon transfer curve method,'' {\em Journal of Innovative
  Optical Health Sciences}~{\bf 9}(03),  1630008 (2016).

\end{thebibliography}
\bibliographystyle{main} 

\end{document}